\documentclass[prd,aps,floats,floatfix,superscriptaddress,preprintnumbers,showpacs,eqsecnum,nofootinbib,twocolumn,showlabels]{revtex4-1}

\usepackage{latexsym,array,theorem,mathrsfs,subfigure,bm,float}

\usepackage[T1]{fontenc}
\usepackage{epsfig}
\usepackage{color}
\usepackage{amsmath}
\usepackage{amsfonts}
\usepackage{amssymb}
\usepackage{graphicx}%
\usepackage{graphics}		
\usepackage{color}

\usepackage[colorlinks=true, a4paper=true, pdfstartview=FitV,linkcolor=blue, citecolor=blue, urlcolor=blue]{hyperref}

\DeclareMathAlphabet{\mathpzc}{OT1}{pzc}{m}{it}

\definecolor{red}{rgb}{1,0,0}
\definecolor{blue }{rgb}{0,0,1}
\definecolor{green}{rgb}{0,1,0}

\usepackage{color}
\input{colordvi.tex}

\newcommand{\be}{\begin{equation}}
\newcommand{\ee}{\end{equation}}
\newcommand{\bea}{\begin{eqnarray}}
\newcommand{\eea}{\end{eqnarray}}

\newcommand{\R}{\overset{\tiny (3)}{R}}
\newcommand{\Rcal}{\overset{\tiny (3)}{\cal R}}

\renewcommand{\theequation}{\arabic{section}.\arabic{equation}}

\begin{document}

\title{Anisotropic universes  in the ghost-free bigravity}

\author{Kei-ichi Maeda}

\affiliation{
Department of Physics, Waseda University, Tokyo 169-8555, Japan
}
\email{maeda@waseda.jp}

\author{Mikhail~S.~Volkov}

\affiliation{
Laboratoire de Math\'{e}matiques et Physique Th\'{e}orique CNRS-UMR 7350, 
Universit\'{e} de Tours, Parc de Grandmont, 37200 Tours, France
}
\email{volkov@lmpt.univ-tours.fr}

\begin{abstract}

We study Bianchi cosmologies in the ghost-free bigravity theory 
assuming both metrics to be homogeneous and anisotropic, of  
the Bianchi class A, which includes types I,II,VI$_0$,VII$_0$,VIII, and IX.
We assume the universe to contain a radiation and a non-relativistic matter,
with the cosmological term mimicked by the graviton mass.  
We find that, for generic initial values leading to a 
late-time self-acceleration, 
the universe approaches a state with non-vanishing anisotropies.
The anisotropy contribution 
to the total energy density decreases much slower than 
in General Relativity and shows the same falloff rate as the energy of 
a non-relativistic matter. 
The solutions show a singularity in the past, and 
in the Bianchi IX case the singularity is approached via a sequence of
Kasner-like steps, which is characteristic for a chaotic behavior. 

\end{abstract}

\maketitle

\section{Introduction}
\setcounter{equation}{0}

The recent discovery of the ghost-free massive gravity theory \cite{deRham:2010kj} 
and its bigravity generalization \cite{Hassan:2011zd} has revived the old idea that 
gravitons can have a small mass \cite{Fierz:1939ix}.
Theories with massive gravitons were for a long time considered as 
pathological,  mainly because they exhibit the Boulware-Deser (BD)
ghost -- an unphysical negative norm state in the spectrum \cite{Boulware:1973my}. 
However, it turns out that the presence of the BD ghost is not
mandatory, and the theories of 
\cite{deRham:2010kj,Hassan:2011zd}
are free of this pathology.
A careful analysis shows  that the number of propagating
degrees of freedom in these theories agrees with the number of graviton
polarizations \cite {Hassan:2011hr,*Golovnev:2011aa,*Kluson:2012wf,*Hassan:2012qv,*Hassan:2011ea}. This does not mean that all solutions are 
stable, since there could be other instabilities,
which should be checked in each particular case. However, since the 
most dangerous BD ghost instability is absent, the ghost-free theories of 
bigravity and massive gravity 
can be considered as healthy  physical models 
for interpreting the observational data.

Theories with massive gravitons can be used in order to 
explain the observed acceleration of our universe 
\cite{1538-3881-116-3-1009,*0004-637X-517-2-565}. 
This effect can be accounted for by introducing a cosmological term to the 
Einstein equations, however, this poses the problem of explaining the origin
and value of this term. An alternative possibility is to consider 
{modifications}
of General Relativity, and theories with massive gravitons are {natural}
candidates for this, since the graviton mass can effectively manifest itself
 as a small cosmological term \cite{PhysRevD.66.104025}. 
This justifies interest towards 
studying cosmological solutions
with massive gravitons. 

The known  cosmologies with massive gravitons can be divided
into two types. The first type is provided by 
solutions described by 
two metrics which are not simultaneously diagonal. 
For these solutions the graviton mass gives rise just to a
constant term in the Einstein equations and to nothing else, at least
for the homogeneous and isotropic backgrounds. Such solutions had been first
 obtained without matter \cite{Koyama:2011xz,*Koyama:2011yg}, 
while later the special 
\cite{Chamseddine:2011bu},\cite{D'Amico:2011jj},\cite{Kobayashi:2012fz,*Gratia:2012wt} 
and also general \cite{Volkov:2012cf,*Volkov:2012zb} 
solutions including a matter source were found. 
For these solutions the mater dominates at early times, when the universe
 is small, while later the mass density decreases and the effective 
cosmological term becomes dominant, leading to a self-acceleration. 
Such solutions exist for all (open, closed and flat) 
Friedmann-Lem\^{a}tre-Robertson-Walker (FLRW) types, both in massive gravity
 and bigravity. However, perturbations around these backgrounds
are expected to be inhomogeneous -- due to the non-diagonal metric components, 
hence solutions of this type are sometimes 
called `inhomogeneous' \cite{D'Amico:2011jj}.

For the second type solutions both metrics are diagonal
and of the FLRW type.  
In this case the effect of the graviton mass 
can be more complex and reduces to that of a cosmological term only 
at late times. Such solutions 
have a somewhat narrow existence 
range. {For example,}
in massive gravity they exist only for the spatially open FLRW type \cite{Gumrukcuoglu:2011ew,*DeFelice:2012mx},
and {although} in bigravity
 they exist for all spatial types  \cite{Volkov:2011an},\cite{vonStrauss:2011mq,*Comelli:2011zm,*Comelli:2012db,*Khosravi:2012rk},
they do not admit the limit where one of the two metrics becomes flat. 
The bigravity solutions exhibit rather complex features and show
several branches. There are physical solutions for which 
the matter dominates at early times,
while the graviton mass becomes essential later. 
In addition, there are also exotic solutions for which the graviton 
mass contribution is dominant at all times.   

Up to now, cosmologies with massive gravitons have been studied mainly in 
the FLRW limit. It may be important to extend the analysis to more 
general spacetimes. For example, 
one may wonder whether the FLRW universe is stable 
against anisotropic or inhomogeneous perturbations \cite{Gumrukcuoglu:2012aa},\cite{Sakakihara:2012iq}.
{Another motivation is the so-called cosmic no-hair conjecture.
In General Relativity (GR) with a cosmological term the
 de Sitter spacetime is expected to be 
an attractor \cite{Starobinsky:1982mr,*Muller:1989rp,*Wald:1983ky,*Kitada:1991ih,*Kitada:1992uh}.
In massive gravity or bigravity the graviton mass gives rise to a 
cosmological term at late times, 
 and so the acceleration of the universe can be explained 
by the de Sitter expansion. However, since the graviton mass is not exactly a 
cosmological term, one may wonder whether the de Sitter space is 
still 
an attractor for generic initial values.}
{An interesting byproduct in such studies can be an observational relic --
if small anisotropy remains in the accelerating phase,
it could be observed \cite{Gumrukcuoglu:2012aa}. }

In what follows, we undertake a systematic analysis of anisotropic 
cosmologies in the ghost-free bigravity \cite{Hassan:2011zd},
assuming both metrics to be simultaneously diagonal and 
of the same Bianchi type 
within the Bianchi class A, 
which includes types  I,II,VI$_0$,VII$_0$,VIII, and IX. 

As a starting point, we find exact solutions for the Bianchi I type for which 
the two metrics have identically the same anisotropies, 
which however requires to fine-tune the two matter sources. 
Then we attack the problem numerically and 
discover that, even in the presence of a matter and for 
all other Bianchi types of class A, 
 the equal anisotropy configurations play 
the role of late time attractors. Specifically, 
starting form arbitrary initial data,
the anisotropies approach equal and constant values.
For Bianchi I solutions constant anisotropies can be scaled away, 
but not  for other Bianchi types,
so that generic homogeneous spacetimes run into anisotropic states.    

We find that the shears  approach zero exponentially fast 
as the universe approaches the 
de Sitter phase. However, 
since the anisotropies do not approach zero but oscillate around constant values, 
the shear contribution to the total energy density 
decreases only as an inverse cube of the size of the universe. 
This is the same 
falloff rate as for a non-relativistic matter, whereas 
in GR shears decrease as the inverse 
sixth power of the size of the universe. 
Therefore, the anisotropy effect could be observable, and so 
 it would be interesting to compare our
predictions with observations. 
Since the anisotropy contribution shows the same falloff rate 
as a cold dark matter, it is tempting to think that the latter could 
in fact be the effect of the 
anisotropies, although it is unclear if 
this interpretation can also explain the 
dark matter clustering. 

The rest of the paper is organized as follows. In the next two sections
we describe the ghost-free bigravity \cite{Hassan:2011zd}, the Bianchi
cosmologies, and derive the field equations in the case where the 
two metrics are simultaneously diagonal and of the same Bianchi type of class A. 
In Section IV we study exact solutions of these equations for the Bianchi I type. 
First, these are solutions with proportional metrics, they are the same as 
in GR. Next, these are solutions with identical anisotropies for the two metrics,
which can be of 
the FLRW type \cite{Volkov:2011an},\cite{vonStrauss:2011mq,*Comelli:2011zm,*Comelli:2012db,*Khosravi:2012rk} 
if anisotropies vanish. 
In Section V we present the generalization to the 
other Bianchi types of class A. We describe our procedure for the numerical 
integration -- 
the dynamical system formulation of the problem and the implementation 
of the initial values constraints, after which we present our numerical results. 
Section VI contains concluding remarks, while in the Appendix we describe 
the Bianchi I solutions in GR and also give the details
of the dynamical system formulation. 

We set $\hbar=c=1$ and use the sign conventions  of Misner-Thorne-Wheeler.
The space and time scale is chosen to be the inverse graviton mass.


\section{The ghost-free bigravity}
\setcounter{equation}{0}
The theory is defined on a four-dimensional spacetime manifold
equipped with two metrics, ${\bf g_{\mu\nu}}$ and 
${\bf f_{\mu\nu}}$. 
The kinetic term of each metric is chosen to be of the 
standard Einstein-Hilbert form, 
while the interaction between them is parametrized by a scalar function 
of the tensor 
\be                             \label{gam}
\gamma^\mu_{~\nu}=\sqrt{{\bf g}^{\mu\alpha}{\bf f}_{\alpha\nu}},
\ee
where ${\bf g}^{\mu\nu}$ is the inverse of ${\bf g}_{\mu\nu}$
and the square root
is understood in the sense that 
\be                     \label{gamgam}
(\gamma^2)^\mu_{~\nu}\equiv \gamma^\mu_{~\alpha}\gamma^\alpha_{~\nu}={\bf g}^{\mu\alpha}
{\bf f}_{\alpha\nu}.
\ee 
The action is
\bea                                      \label{1}
&&S[{\bf g},{\bf f},{\rm matter}]
\nonumber \\
&&~~=\frac{1}{2\kappa_g^2}\,\int d^4x\, \sqrt{-{\bf g}}\,R({\bf g})
+\frac{1}{2\kappa_f^2}\,\int d^4x\, \sqrt{-{\bf f}}\, {\cal R}({\bf f})  \notag \\
&&~~-\frac{m^2}{\kappa^2}\int d^4x\, \sqrt{-{\bf g}} \, \mathscr{U}[{\bf g},{\bf f}] \,
 \\
&&~~+S^{[\rm m]}_g[{\bf g},{\rm g\mathchar`-matter}]
+S^{[\rm m]}_f[{\bf f},{\rm f\mathchar`-matter}] \nonumber
\,, 
\eea
where $R$ and ${\cal R}$ are the Ricci scalars for ${\bf g}_{\mu\nu}$ and 
${\bf f}_{\mu\nu}$, respectively,  $\kappa_g^2=8\pi G$ and $\kappa_f^2=8\pi {\cal G}$
 are the corresponding gravitational couplings, while 
$\kappa^2=\kappa_g^2+\kappa_f^2$ and 
$m$ is the graviton mass.
The interaction 
between the two metrics is given by
\be                             \label{2}
\mathscr{U}=\sum_{k=0}^4 b_k\,\mathscr{U}_k(\gamma),
\ee
where $b_k$ are parameters, while  
$\mathscr{U}_k(\gamma)$ are defined 
by the following 
relations\footnote{Notice that $\epsilon^{0123}=-\epsilon_{0123}=1$.} 
\bea                        \label{4}
\mathscr{U}_0(\gamma)&=&
-\frac{1}{4!}\,\epsilon_{\mu\nu\rho\sigma}\epsilon^{\mu\nu\rho\sigma} =1,  \\
\mathscr{U}_1(\gamma)&=&-\frac{1}{3!}\,
\epsilon_{\mu\nu\rho\sigma}\epsilon^{\alpha\nu\rho\sigma}\gamma^\mu_{~\alpha}
=\sum_{A}\lambda_A=[\gamma],\notag \\
\mathscr{U}_2(\gamma)&=&-\frac{1}{2!}\,
\epsilon_{\mu\nu\rho\sigma}\epsilon^{\alpha\beta\rho\sigma}
\gamma^\mu_{~\alpha}\gamma^\nu_{~\beta}  
=\sum_{A<B}\lambda_A\lambda_B \notag \\
&=&\frac{1}{2!}([\gamma]^2-[\gamma^2]),\notag \\
\mathscr{U}_3(\gamma)&=&-\frac{1}{3!}\,
\epsilon_{\mu\nu\rho\sigma}\epsilon^{\alpha\beta\gamma\sigma}
\gamma^\mu_{~\alpha}\gamma^\nu_{~\beta}\gamma^\rho_{~\gamma}
=\sum_{A<B<C}\lambda_A\lambda_B\lambda_C
\nonumber \\
&=&
\frac{1}{3!}([\gamma]^3-3[\gamma][\gamma^2]+2[\gamma^3]),\notag \\
\mathscr{U}_4(\gamma)&=&-\frac{1}{4!}\,
\epsilon_{\mu\nu\rho\sigma}\epsilon^{\alpha\beta\gamma\delta}
\gamma^\mu_{~\alpha}\gamma^\nu_{~\beta}
\gamma^\rho_{~\gamma}\gamma^\sigma_{~\delta}
=
\lambda_0\lambda_1\lambda_2\lambda_3
\nonumber \\
&=&
\frac{1}{4!}([\gamma]^4-6[\gamma]^2[\gamma^2]+8[\gamma][\gamma^3]+3[\gamma^2]^2
-6[\gamma^4])\,. \notag 
\eea
Here $\lambda_A$ ($A=0,1,2,3$) are eigenvalues of $\gamma^\mu_{~\nu}$, 
and, using the hat to denote matrices, we have defined 
\be                       \label{bracket}
[\gamma]={\rm tr}(\hat{\gamma})=\gamma^\mu_{~\mu},~~~~
[\gamma^k]={\rm tr}(\hat{\gamma}^k)=(\gamma^k)^\mu_{~\mu}
\,. 
\ee
 The above choice of the interaction
potential $\mathscr{U}$ insures that the BS ghost is absent 
\cite {Hassan:2011hr,*Golovnev:2011aa,*Kluson:2012wf,%
*Hassan:2012qv,*Hassan:2011ea}.
We have also assumed a g-matter and an f-matter 
interacting, respectively, only with ${\bf g}_{\mu\nu}$ 
and with ${\bf f}_{\mu\nu}$. One cannot have a matter coupled
to both metrics at the same time, since  the BD ghost would 
come back in this case, whereas our choice 
is ghost-free \cite{Hassan:2011zd}. In addition, this choice preserves 
the equivalence principle, since the g-matter follows 
geodesics of the g-metric and the f-matter follows f-geodesics.  
Although it is sometimes  convenient to have only a g-matter and 
choose the f-sector to be empty, nothing forbids to have both matter types
at the same time.

It is also useful to express the interaction in terms of 
${\cal K}^\mu_\nu=\delta^\mu_\nu-\gamma^\mu_{~\nu}$, 
\be                             \label{2a}
\mathscr{U}=\sum_{k=0}^4 c_k\,\mathscr{U}_k({\cal K}), 
\ee
where $\mathscr{U}_k({\cal K})$ are defined by the same expressions as
 in \eqref{4}, up to the replacement  
$\lambda_A\to \mu_A=1-\lambda_A$ and $\gamma\to {\cal K}$, where
$\mu_A$'s are eigenvalues of ${\cal K}^\mu_{~\nu}$. 
The parameters  $c_k$ are related to $b_k$ as 
\bea
c_0&=&b_0+4b_1+6b_2+4b_3+b_4, \notag \\
c_1&=&-(b_1+3b_2+3b_3+b_4),\notag \\
c_2&=& b_2+2b_3+b_4,\notag \\
c_3&=& -(b_3+b_4),\notag \\
c_4&=&b_4\,,
\eea
with the inverse expressions
\bea
b_0&=&c_0+4c_1+6c_2+4c_3+c_4,\notag \\
b_1&=&-(c_1+3c_2+3c_3+c_4),\notag \\
b_2&=&c_2+2c_3+c_4,   \notag \\
b_3&=&-(c_3+c_4), \notag \\
b_4&=&c_4.
\eea
In the weak field limit, when 
${\bf g}_{\mu\nu}\approx {\bf f}_{\mu\nu}\approx \eta_{\mu\nu}$, 
the tensor ${\cal K}$ tends to zero while 
$\mathscr{U}_k\sim {\cal K}^k$ so that 
Eq.\eqref{2a} gives the interaction in terms 
of powers of deviation from the flat space. 
We require the flat space to be a solution of the field
equations, which is only possible if 
\be                                            \label{a2}
c_0=c_1=0,
\ee
while the quadratic part of the interaction should 
reproduce the Fierz-Pauli term,
\be
+\frac{m^2}{2}([{\cal K}]^2-[{\cal K}^2]),
\ee
which fixes the normalization 
\be                                         \label{a1}
c_2=-1.
\ee  
The above choice  implies that the bare cosmological constant is zero,
so that a non-zero cosmological term can only be of a dynamical origin, 
due to the graviton mass contribution.   
Terms proportional to $c_3$ and$ 
c_4$ can be kept, so that 
\bea                \label{bbb}
&&
b_0=4c_3+c_4-6,~~
b_1=3-3c_3-c_4,~~
\nonumber \\
&&
b_2=2c_3+c_4-1,~~
b_3=-(c_3+c_4),~~
\nonumber \\
&&
b_4=c_4.
\eea

\subsection{Field equations}
Assuming the spacetime coordinates $x^\mu$ to be dimensionless, 
the metrics ${\bf g}_{\mu\nu}$, ${\bf f}_{\mu\nu}$ have the 
dimension of length squared. 
To pass to dimensionless quantities, we make a conformal rescaling 
\be
{\bf g}_{\mu\nu}=\frac{1}{m^2}\,g_{\mu\nu},~~~~~
{\bf f}_{\mu\nu}=\frac{1}{m^2}\,f_{\mu\nu}. 
\ee
Varying the action then gives the field equations 
\bea                                  \label{Einstein}
G_{\mu\nu}&=&\kappa_g^2 \left[T_{\mu\nu}^{[\gamma]}
+\frac{1}{m^2}\,\underline{T}^{\rm [m]}_{~\mu\nu}\right]
\,, \\
{\cal G}_{\mu\nu}&=&\kappa_f^2\left[{\cal T}_{\mu\nu}^{[\gamma]}+ 
\frac{1}{m^2}\,\underline{\cal T}^{\rm [m]}_{~\mu\nu}\right]\,, \label{Einstein1}
\eea
where $G_{\mu\nu}$ and ${\cal G}_{\mu\nu}$ are the  Einstein tensors for $g_{\mu\nu}$
and $f_{\mu\nu}$. 
The graviton energy-momentum tensors are 
\bea                      \label{var}
&&
T_{\mu\nu}^{[\gamma]}=\frac{1}{\kappa^2}
\left(2\,\frac{\delta\mathscr{U}}{\delta g^{\mu\nu}}
-\mathscr{U}\,g_{\mu\nu}
\right),\nonumber \\
&&
{\cal T}_{\mu\nu}^{[\gamma]}=2\,\frac{1}{\kappa^2}
\frac{\sqrt{-g}}{\sqrt{-f}}\,\frac{\delta\mathscr{U}}{\delta f^{\mu\nu}}\,.
\eea
In order to perform the variations here, one uses 
the relations  
\bea
&&
\frac{\delta[\gamma^n]}{\delta g^{\mu\nu}}=\frac{n}{2}\,
g_{\mu\alpha}(\gamma^n)^\alpha_{~\nu}=\frac{n}{2}\,
g_{\nu\alpha}(\gamma^n)^\alpha_{~\mu},\nonumber \\
&&
\frac{\delta[\gamma^n]}{\delta f^{\mu\nu}}=-\frac{n}{2}\,
f_{\mu\alpha}(\gamma^n)^\alpha_{~\nu}=-\frac{n}{2}\,
f_{\nu\alpha}(\gamma^n)^\alpha_{~\mu}
\,,
\eea
which can be obtained by varying the 
definition $\gamma^\mu_{~\nu}=\sqrt{{g}^{\mu\alpha}{f}_{\alpha\nu}}$ 
and using the properties of the trace. 
Introducing the angle $\eta$ such that
\be
\kappa_g=\kappa\cos\eta,~~~~~~\kappa_f=\kappa\sin\eta,
\ee
one obtains dimensionless quantities 
\bea                        \label{T}
&&
\kappa_g^2 T^{[\gamma]\mu}_{~~~~~\nu}=
\cos^2\eta(\,\tau^\mu_{~\nu}-\mathscr{U}\,\delta^\mu_\nu),
\nonumber \\
&&
\kappa_f^2 {\cal T}^{[\gamma]\mu}_{~~~~~\nu}
=-\sin^2\eta\,\frac{\sqrt{-g}}{\sqrt{-f}}\,\tau^\mu_{~\nu}\,,
\eea
where $T^\mu_{~\nu}=g^{\mu\alpha}T_{\alpha\nu}$ and 
${\cal T}^\mu_{~\nu}=f^{\mu\alpha}{\cal T}_{\alpha\nu}$
while
\bea                                \label{tau}
\tau^\mu_{~\nu}&=&
\{b_1\,\mathscr{U}_0+b_2\,\mathscr{U}_1+b_3\,\mathscr{U}_2
+b_4\,\mathscr{U}_3\}\gamma^\mu_{~\nu} \notag 
\nonumber \\
&-&\{b_2\,\mathscr{U}_0+b_3\,\mathscr{U}_1+b_4\,\mathscr{U}_2\}(\gamma^2)^\mu_{~\nu}  \\
&+&\{b_3\,\mathscr{U}_0+b_4\,\mathscr{U}_1\}(\gamma^3)^\mu_{~\nu} \notag  \\
&-&b_4\,\mathscr{U}_0\,(\gamma^4)^\mu_{~\nu}
\eea
with $\mathscr{U}_k\equiv \mathscr{U}_k(\gamma)$. 
The matter energy-momentum tensors $\underline{T}^{\rm [m]}_{~\mu\nu}$
and $\underline{\cal T}^{\rm [m]}_{~\mu\nu}$ are dimensionful, we 
assume them to be  of perfect fluid type. However, they enter the equations
only via dimensionless combinations%
\footnote{We include the gravitational constants $\kappa_g^2$ and $\kappa_f^2$
and the graviton mass $m$ 
in the definition of the energy densities and pressures.} 
\bea                        \label{Tm}
&&
\frac{\kappa_g^2}{m^2}\, \underline{T}^{\rm [m]\mu}_{~~~~~\nu}\equiv T^{\rm [m]\mu}_{~~~~~\nu}=
(\rho_g+P_g){U}^\mu {U}_\nu+P_g\,
\delta^\mu_\nu
\nonumber \\
&&
\frac{\kappa_f^2}{m^2}\, \underline{\cal T}^{\rm [m]\mu}_{~~~~~\nu}
\equiv {\cal T}^{\rm [m]\mu}_{~~~~~\nu}=
(\rho_f+P_f){\cal U}^\mu {\cal U}_\nu+P_f\,
\delta^\mu_\nu
\,.~~~~~~~~
\eea
As a result, from now on we shall be considering the field equations 
\eqref{Einstein},\eqref{Einstein1} 
expressed entirely in terms of dimensionless quantities. 

The diffeomorphism invariance of the matter terms in the action 
implies the 
conservation conditions
\bea
\stackrel{(g)}{\nabla}_\mu T^{\rm [m]\mu}_{~~~~~\nu}=0,~~~~~
\stackrel{(f)}{\nabla}_\mu {\cal T}^{\rm [m]\mu}_{~~~~~\nu}=0
\,,
\label{matter_cons}
\eea
where $\stackrel{(g)}{\nabla}$ and $\stackrel{(f)}{\nabla}$
are covariant derivatives with respect to $g_{\mu\nu}$ and 
$f_{\mu\nu}$. 
The Bianchi identities for \eqref{Einstein} then imply that  
\be                                   \label{Bian} 
\stackrel{(g)}{\nabla}_\mu\! T^{[\gamma]\mu}_{~~~~~\nu}=0. 
\ee
Similarly, the Bianchi identities for equations \eqref{Einstein1} imply that 
$\stackrel{(f)}{\nabla}_\mu\!{\cal T}^{[\gamma]\mu}_{~~~~~\nu}
=0$, but in fact 
this condition is not independent and follows from \eqref{Bian} 
 in view of the diffeomorphism invariance of 
the interaction term 
in the action.

\section{Bianchi Spacetimes}
\setcounter{equation}{0}

In what follows, we shall assume 
both metrics to be homogeneous but anisotropic, that is, 
invariant  under a three-parameter
translation group G$_3$ acting on the 3-space. 
The group is generated by three vector fields $e_a$  
satisfying commutation relations 
\be
[e_a,e_b]=C^c_{~ab}e_c
\ee
with constant structure coefficients $C^c_{~ab}$. 
Such groups have been all classified by Bianchi. 
The Jacobi identities imply that 
structure coefficients can be parameterized as 
\be                           \label{C}
C^c_{~ab}=n^{cd}\epsilon_{dab}+a(\delta^1_a\delta_b^c-\delta^1_b\delta_a^c)
\,
\ee
with $n^{ab}={\rm diag}[n^{(1)},n^{(2)},n^{(3)}]$.
The different choices of the four parameters $n^{(1)}$, $n^{(2)}$, $n^{(3)}$,
 $a$
correspond to the nine Bianchi types.

Denoting $\omega^a$ the 1-forms dual to $e_a$, 
the two metrics can be parameterized as  
\bea                                                    \label{BianA}
ds_g^2&=&-\alpha(t)^2dt^2+h_{ab}(t)\,\omega^a\otimes \omega^b, \notag \\
ds_f^2&=&-{\mathcal A}(t)^2dt^2+{\cal H}_{ab}(t)\,\omega^a\otimes\omega^b\,.
\eea
In this paper, we discuss only the class A Bianchi models,
for which the parameter $a$ in \eqref{C} vanishes. These 
include types I, II, VI$_0$, VII$_0$, VIII, IX.
For these types tensors $h_{ab}$ and ${\cal H}_{ab}$ can be chosen to 
be diagonal, and this guarantees that $G_{0a}=0$ and ${\cal G}_{0a}=0$,
so that there are no energy fluxes. 
To achieve a similar no-flux condition for the type B (tilted) 
Bianchi classes with $a\neq 0$, one has to either let 
$h_{ab}$ and ${\cal H}_{ab}$ be non-diagonal,
or to impose constraints 
on the otherwise independent values of their components. 

Let $h_{ab}$, ${\cal H}_{ab}$ be diagonal matrices,
\begin{eqnarray*}
h_{ab}&=&{\rm diag}[\alpha_1^{\,2},\alpha_2^{\,2},\alpha_3^{\,2}],
\\
{\cal H}_{ab}&=&
{\rm diag}[{\cal A}_1^{\,2},{\cal A}_2^{\,2},{\cal A}_3^{\,2}],
\end{eqnarray*}
and introduce 
\begin{eqnarray*}             \label{pm1}
Q^a_{~b}&=&2\,{\rm diag}\left[
\frac{\dot{\alpha_1}}{\alpha_1},
\frac{\dot{\alpha_2}}{\alpha_3},
\frac{\dot{\alpha_3}}{\alpha_3}\right]
\,,
\\
{\cal Q}^a_{~b}&=&2\,{\rm diag}\left[
\frac{\dot{{\cal A}_1}}{{\cal A}_1},
\frac{\dot{{\cal A}_2}}{{\cal A}_3},
\frac{\dot{{\cal A}_3}}{{\cal A}_3}\right]
\,,~~
\end{eqnarray*}
where the dot denotes the time derivative. The non-zero projections 
$R^A_{~B}$ and ${\cal R}^A_{~B}$ 
of the Ricci tensors on the tetrad base $\Theta^A=(dt,\omega^a)$ and 
$E_A=(\partial_t,e_a)$ read 
\begin{eqnarray}
&&R^0_{~0}=\frac{[\dot{Q}]}{2}+\frac{[Q^2]}{4},~~
R^a_{~b}=\,\frac{\left(\sqrt{h}Q^a_{~b}\right)^\centerdot}{2\alpha\sqrt{h}}
+\overset{(3)~}{R^a_{~b}},~ \notag \\
&&{\cal R}^0_{~0}=\frac{[\dot{\cal Q}]}{2}+\frac{[{\cal Q}^2]}{4},~~
{\cal R}^a_{~b}=\frac{\left(\sqrt{\cal H}{\cal Q}^a_{~b}\right)^\centerdot}
{2\sqrt{\cal H}{\cal A}}
+\overset{(3)~}{{\cal R}^a_{~b}}.~~~\label{RRRR}
\end{eqnarray}
Here the bracketed quantities are calculated according to \eqref{bracket},
while the three-dimensional Ricci tensors  
\bea
h\,\overset{(3)~}{R^a_{~b}}=2(N^2)^a_{~b}
-[N]\,N^a_{~b}
-([N^2]-{1\over 2}[N]^2)\delta^a_{~b}
\,,~~~~\notag \\
{\cal H}\,\overset{(3)~}{{\cal R}^a_{~b}}=2({\cal N}^2)^a_{~b}
-[{\cal N}]\,{\cal N}^a_{~b}
-([{\cal N}^2]-{1\over 2}[{\cal N}]^2)\delta^a_{~b}
\,,~~~~
\eea
where 
\be
N^a_{~b}=n^{ac}h_{cb},~~~~ 
{\cal N}^a_{~b}=n^{ac}{\cal H}_{cb}.
\ee 
The Ricci scalars are
\be
\R=\frac{1}{2h}\left([N]^2-2[N^2]\right),~~
\Rcal=\frac{1}{2{\cal H}}\left([{\cal N}]^2-2[{\cal N}^2]\right).~~
\ee

The non-trivial tetrad projections $\gamma^A_{~B}$ of the tensor 
$\gamma^\mu_{~\nu}=\sqrt{g^{\mu\alpha}f_{\alpha\nu}}$ 
are 
\be                              \label{gm}
\gamma^A_{~B}={\rm diag}\left[
\frac{{\cal A}}{\alpha},
\frac{{\cal A}_1}{\alpha_1},
\frac{{\cal A}_2}{\alpha_2},
\frac{{\cal A}_3}{\alpha_3}
\right].
\ee
Using this in Eqs.\eqref{2},\eqref{4},\eqref{T},\eqref{tau}
reveals that non-trivial tetrad components of the 
energy-momentum tensors $T^{[\gamma]\mu}_{~~~~~\nu}$
and ${\cal T}^{[\gamma]\mu}_{~~~~~\nu}$ are also diagonal.
For example, the non-trivial tetrad projections of  
$\tau^\mu_{~\nu}$ defined in \eqref{tau} are
\begin{eqnarray*}                          
&&               \label{tau1}
\tau^A_{~A}({\rm no~sum})=
\notag \\
&&
\lambda_A\left[b_1+b_2\sum_{B\neq A}\lambda_B
+b_3\hskip -.5cm \sum_{B,C\neq A;B<C}\hskip -.5cm \lambda_B\lambda_C \right]
+b_4\lambda_0\lambda_1\lambda_2\lambda_3, 
\end{eqnarray*}
which defines the non-trivial components 
\begin{eqnarray*}
T^{[\gamma]A}_{~~~~A}({\rm no~sum})
&=&\frac{1}{\kappa^2}\left[\tau^A_{~A}({\rm no~sum})-\mathscr{U}
\right]\,,
\\
{\cal T}^{[\gamma]A}_{~~~~A}({\rm no~sum})&=&
-\frac{1}{\kappa^2} 
\frac{{\cal A}\sqrt{\cal H}}{\alpha\sqrt{h}}\,\tau^A_{~A}({\rm no~sum})\,,
\end{eqnarray*}
with
$$\mathscr{U}=b_0+b_1\sum_B \lambda_B+\ldots 
+b_4\,\lambda_0\lambda_1\lambda_2\lambda_3\,.
$$
This implies that
\be
\int \mathscr{U}\sqrt{-g}\,d^4x=\int (\alpha U_g+{\cal A}\,{\cal U}_f)d^4x
\ee
where 
\bea
U_g&=&
\sqrt{h}\,\{b_0+b_1(\lambda_1+\lambda_2+\lambda_3)\\
&+&b_2(\lambda_1\lambda_2
+\lambda_2\lambda_3+\lambda_3\lambda_1)
+b_3\lambda_1\lambda_2\lambda_3\} 
\eea
and ${\cal U}_f$ is obtained from this by replacing 
$h\to {\cal H}$ and $b_k\to b_{k+1}$.

Finally, the matter energy-momentum tensors \eqref{Tm},
assuming the fluid four-velocities to be tangent to the timelines 
and $\rho_g$, $P_g$, $\rho_f$, $P_f$ to depend only on time, are
\bea                                 \label{TTm}
T^{[m]A}_{~~~~~B}&=&{\rm diag}[-\rho_g,P_g,P_g,P_g],~\notag \\
{\cal T}^{[m]A}_{~~~~~B}&=&{\rm diag}[-\rho_f,P_f,P_f,P_f].~~
\eea
Inserting the above expressions \eqref{RRRR}--\eqref{TTm} 
to the Einstein equations \eqref{Einstein},\eqref{Einstein1} 
gives a system of non-linear ordinary differential \
equations for the field amplitudes $\alpha$, $ {\cal A}$, 
$\alpha_a$, ${\cal A}_{a}$,
 $\rho_g$, $P_g$, $\rho_f$, $P_f$ which depend only on time. 

It is instructive to rederive these equations  
in a different way, by first inserting the metrics \eqref{BianA} 
to the action and then varying with respect to the field amplitudes. 
\begin{widetext}
We adopt the following parametrization:
\be            \label{pm2}
[\alpha_1,\alpha_2,\alpha_3]=e^{\Omega}\,\times\left[
e^{\beta_{+}+\sqrt{3}\beta_{-}} , e^{\beta_{+}-\sqrt{3}\beta_{-}},
 e^{-2\beta_{+}} 
\right]
\,,~~
\left[
{\cal A}_1,
{\cal A}_2,
{\cal A}_3
\right]
=
e^{\cal W}\,\times\left[
e^{{\cal B}_{+}+\sqrt{3}{\cal B}_{-}},
e^{ {\cal B}_{+}-\sqrt{3}{\cal B}_{-}} ,
e^{-2{\cal B}_{+}}\right]
\,. 
\ee
The action \eqref{1} then assumes the form 
\bea
m^2S&=&\frac{1}{2\kappa_g^2}\,
\int \, \alpha e^{3\Omega}\left\{{6\over \alpha^2}\left(-\dot \Omega^2
+\dot \beta_+^2+\dot \beta_-^2\right)+\R\right\}d^4x
+\frac{1}{2\kappa_f^2}\,\int\, \,{\cal A}e^{3{\cal W}}
\left\{
{6\over {\cal A}^2}\left(-\dot {\cal W}^2
+\dot {\cal B}_+^2+\dot {\cal B}_-^2\right) 
+\Rcal\right\}d^4x \notag \\
&-&\frac{1}{\kappa^2}\int \,  \, \{\alpha U_g+ {\cal A}\,{\cal U}_f  \}d^4x +
\frac{1}{\kappa_g^2}S_g^{\rm [m]}[g,\rho_g,P_g]+
\frac{1}{\kappa_f^2}{\cal S}_f^{\rm [m]}[f,\rho_f,P_f]
\,, 
\eea
where 
\bea
U_g&=&
b_0 e^{3\Omega}+b_3 e^{3{\cal W}}+b_1e^{{\cal W}+2\Omega}
\left(e^{-2({\cal B}_+-\beta_+)}+
2e^{{\cal B}_+-\beta_+}\cosh[\sqrt{3}({\cal B}_--\beta_-)]
\right)
\nonumber \\
&&
~~~~+b_2 e^{2{\cal W}+\Omega}\left(e^{2({\cal B}_+-\beta_+)}+
2e^{-({\cal B}_+-\beta_+)}\cosh[\sqrt{3}({\cal B}_--\beta_-)]\right),
\eea
while ${\cal U}_f$ is obtained from this by replacing 
$b_0\to b_{1}$,
$b_1\to b_{2}$,
$b_2\to b_{3}$,
$b_3\to b_{4}$,
whereas 
\bea
\R&=&
2n^{(1)}n^{(3)}e^{-2\Omega}e^{-2(\beta_{+}-\sqrt{3}\beta_{-})}
-\frac12\,e^{-2\Omega}\left\{
n^{(1)}e^{2(\beta_{+}+\sqrt{3}\beta_{-})}
-n^{(2)}e^{2(\beta_{+}-\sqrt{3}\beta_{-})}
+n^{(3)}e^{-4\beta_{+}}\right\}^2
\,,\label{R}
\eea
and $\Rcal$ is obtained from this by replacing $\Omega\to{\cal W}$
and $\beta_\pm\to{\cal B}_\pm$.  

Varying the action with respect to 
$\alpha,\Omega,\beta_\pm$
yields the equations 
\bea
\left(e^{3\Omega}\,\frac{\dot{\Omega}}{\alpha}\right)^2&=&
\left(e^{3\Omega}\,\frac{\dot{\beta}_{+}}{\alpha}\right)^2
+\left(e^{3\Omega}\,\frac{\dot{\beta}_{-}}{\alpha}\right)^2
+{1\over 6}\left[2\cos^2\eta\,  e^{3\Omega}U_g
-\,e^{6\Omega}\R
+2e^{6\Omega}\rho_g\right]\,,\label{C1} \\
\left(e^{3\Omega}\,\frac{\dot{\Omega}}{\alpha}\right)^{\centerdot}
&=&{1\over 6}\left[
\cos^2\eta  \left(\frac{\partial U}{\partial \Omega}+3\alpha\,{\cal U}_g\right)
-2\alpha \,e^{3\Omega}\R
+3 \alpha \,e^{3\Omega}(\rho_g-P_g)\right]\,, \label{e1}\\
\left(e^{3\Omega}\,\frac{\dot{\beta}_{\pm}}{\alpha}\right)^{\centerdot}
&=&-{1\over 12}
\frac{\partial}{\partial{\beta_\pm}} \left(2 \cos^2\eta\,  U-{\alpha}\,
e^{3\Omega}\R\right),              \label{e2}
\eea
while varying with respect to 
${\cal A},{\cal W},{\cal B}_\pm$ one obtains 
\bea
\left(e^{3{\cal W}}\,\frac{\dot{{\cal W}}}{{\cal A}}\right)^2&=&
\left(e^{3{\cal W}}\,\frac{\dot{\cal B}_{+}}{{\cal A}}\right)^2
+\left(e^{3{\cal W}}\,\frac{\dot{\cal B}_{-}}{{\cal A}}\right)^2
+{1\over 6}\left[2 \sin^2\eta\, e^{3{\cal W}}{\cal U}_f
-\,e^{6{\cal W}}\Rcal
+2e^{6{\cal W}}\rho_f\right]
\,,\label{C2} \\
\left(e^{3{\cal W}}\,\frac{\dot{{\cal W}}}{\cal A}\right)^{\centerdot}
&=&{1\over 6}\left[
\sin^2\eta \left(\frac{\partial U}{\partial{\cal W}}+3{\cal A}{\cal U}_f\right)
-2 {\cal A}\,e^{3\cal W}\Rcal
+3{\cal A}\,e^{3{\cal W}}\left(\rho_f-P_f\right)\right]\,,
         \label{e3}\\
\left(e^{3{\cal W}}\,\frac{\dot{B}_{\pm}}{\cal A}\right)^{\centerdot}
&=&-{1\over 12}
\frac{\partial}{\partial{{\cal B}_\pm}} \left(2 \sin^2\eta\, U-{\cal A}\,
e^{3\cal W}\Rcal\right)\,.  \label{e4}
\eea
Here 
$U=\alpha U_g+{\cal A}\,{\cal U}_f$.
The matter densities and pressures satisfy the 
conservation conditions 
\be                             \label{mat}
\dot{\rho}_g+3\dot{\Omega}(\rho_g+P_g)=0,~~~~~
\dot{\rho}_f+3\dot{{\cal W}}(\rho_f+P_f)=0. 
\ee
Differentiating the first order constraint \eqref{C1} and using
 \eqref{e1},\eqref{e2}
to eliminate the second derivatives gives
\be                              \label{C:a}
\alpha\,\left(\dot{W}\frac{\partial}{\partial{\cal {\cal W}}}
+\dot{\cal B}_{+}\frac{\partial}{\partial{\cal {\cal B}_{+}}}
+\dot{\cal B}_{-}\frac{\partial}{\partial{\cal {\cal B}_{-}}}\right)U_g=
{\cal A}\,\left(\dot{\Omega}\frac{\partial}{\partial\Omega}
+\dot{\beta}_{+}\frac{\partial}{\partial{\beta_{+}}}
+\dot{\beta}_{-}\frac{\partial}{\partial{\beta_{-}}}
\right){\cal U}_f\,,
\ee
which is nothing but the condition of conservation of 
$T^{[\gamma]\mu}_{~~~~\nu}$. 
Differentiating the first order constraint \eqref{C2} and using
 \eqref{e3},\eqref{e4}
to eliminate the second derivatives reproduces the same condition again, 
because ${\cal T}^{[\gamma]\mu}_{~~~~\nu}$
 is automatically conserved as soon as  
$T^{[\gamma]\mu}_{~~~~\nu}$ is conserved. 

Since the structure of the two line elements in \eqref{BianA} is 
invariant under time reparametrizations, a gauge condition 
can be imposed. For example, one can fix the gauge by requiring that 
$\alpha=1$.

Equations \eqref{C1}--\eqref{e4}
are equivalent to those obtained in the component approach by
inserting \eqref{pm1}--\eqref{TTm} 
to \eqref{Einstein},\eqref{Einstein1}. 
The equivalence is seen in view of the relations 
\bea
2\overset{(3)~}{R^3_{~3}}-
\overset{(3)~}{R^2_{~2}}-
\overset{(3)~}{R^1_{~1}}&=&\frac12\frac{\partial \R}{\partial \beta_{+}},~~~~~~
\R=-\frac12\frac{\partial \R}{\partial\Omega},~~~
~~
\overset{(3)~}{R^2_{~2}}-
\overset{(3)~}{R^1_{~1}}=\frac{1}{2\sqrt{3}}\frac{\partial \R}{\partial \beta_{-}},
\eea
and 
\bea                           \label{TT}
&&\kappa^2e^{3\Omega}\,T^{[\gamma]0}_{~~~~0}=-U_g\,,~~~
\kappa^2\alpha\,e^{3\Omega}\,(3T^{[\gamma]0}_{~~~~0}+T^{[\gamma]1}_{~~~~1}
+T^{[\gamma]2}_{~~~~2}+T^{[\gamma]2}_{~~~~2})=
-\left(\frac{\partial U}{\partial\Omega}
+3\alpha U_g\right)\,,~~~ \notag \\
&&\kappa^2\alpha\,e^{3\Omega}\,(2T^{[\gamma]3}_{~~~~3}
-T^{[\gamma]2}_{~~~~2}-T^{[\gamma]1}_{~~~~1})=
\frac{\partial U}{\partial\beta_{+}}\,,~~~~
\sqrt{3}\,\kappa^2\alpha\,e^{3\Omega}\,(T^{[\gamma]2}_{~~~~2}-T^{[\gamma]1}_{~~~~1})=
\frac{\partial U}{\partial\beta_{-}}\,,
\eea
as well as those obtained from these by replacing 
$\overset{(3)~}{R^a_{~b}}\to \overset{(3)~}{{\cal R}^a_{~b}}$, 
$T^{[\gamma]\mu}_{~~~~\nu}\to {\cal T}^{[\gamma]\mu}_{~~~~\nu}$,  
$U_g\to {\cal U}_f$, $\alpha\to {\cal A}$, $\Omega\to {\cal W}$, 
$\beta_\pm\to{\cal B}_\pm$.  

Assuming the matter to consist of several components labeled by $i$
with pressure being proportional to the energy density for each component
(for example radiation plus a non-relativistic component),  
the matter densities and pressures determined by \eqref{mat} are 
(with constant $\rho_{g}^{(i)}$ and $w^{(i)}_g$) 
\be                              \label{mat1}
\rho_g=\sum_j \rho_{g}^{(i)}e^{-3(1+w^{(i)}_g)\Omega},~~~
P_g=\sum_i w^{(i)}_g\rho_{g}^{(i)}e^{-3(1+w^{(i)}_g)\Omega}. 
\ee
Similar expressions for $\rho_f,P_g$ are obtained by replacing the index $g\to f$
and $\Omega\to{\cal W}$. 

\vspace{2 mm}

\end{widetext}

\section{Bianchi Type I solutions}

\setcounter{equation}{0}

Let us first discuss the simplest Bianchi type I model, in which case the 
isometry 
group G$_{3}$ acting on the 3-space is abelian.
The two metrics are 
\bea
ds_g^2=&-&\alpha(t)^2dt^2+ \sum_{a=1,2,3}\alpha_a^2(t)(dx^a)^2\,,\notag \\
ds_f^2=&-&{\cal A}(t)^2dt^2+ \sum_{a=1,2,3}{\cal A}^2_a(t)(dx^a)^2\,,
\eea
with $\alpha_a,{\cal A}_a$ parameterized according to \eqref{pm2}. 
Since the spatial  curvatures vanish, 
$$
\overset{(3)~}{R^a_{~b}}=\overset{(3)~}{{\cal R}^a_{~b}}=0\,,
$$
the basic equations simplify, which allows us to find
some exact solutions. 

\subsection{Recovering General Relativity}
\label{GR}

If we assume the two
metrics to be  proportional, 
\be                          \label{prop}
f_{\mu\nu}=C^2 g_{\mu\nu},
\ee
then we obtain the 
GR solutions. Indeed, one has in this case 
$\gamma^\mu_{~\nu}=C\delta^\mu_{~\nu}$ and so 
\bea
\tau^\mu_{~\nu}&=&
(b_1 +3 b_2\,C+3 b_3\,C^2
+ b_4\,C^3)C \delta^\mu_{~\nu}
\,,
\eea
which gives the energy-momentum tensors 
\bea
\kappa_g^2 T^{[\gamma]\mu}_{~~~~\nu}&=&-\Lambda_g(C) \delta^\mu_{~\nu}\,,~\notag \\
\kappa_f^2 {\cal T}^{[\gamma]\mu}_{~~~~\nu}&=&-\Lambda_f(C) \delta^\mu_{~\nu}
\,,
\eea
with 
\begin{eqnarray}
\Lambda_g(C) &=&\cos^2\eta
\left(b_0 +3 b_1\,C+3 b_2\,C^2
+ b_3\,C^3\right)\,,~
\label{Lmbd}  \notag 
\\
\Lambda_f(C) &=&{\sin^2\eta\over C^3}
\left(b_1 +3 b_2 C\,+3 b_3 C^2\,
+ b_4 C^3\,\right)\,.
\end{eqnarray}
Since  the energy-momentum tensors should be conserved, 
it follows that $C$ is a constant.
As a result, 
we find two sets of Einstein equations:
\bea      
&&
G_\mu^\nu   +\Lambda_g \delta_\mu^\nu=
T^{[{\rm m}]\mu}_{~~~~\nu} \,,    \label{GRa}
\\
&&                                \label{GRb}
{\cal G}_\mu^\nu    +\Lambda_f \delta_\mu^\nu=
{\cal T}^{[{\rm m}]\mu}_{~~~~\nu} 
\,.
\eea
Since one has 
${\cal G}_\mu^\nu=  G_\mu^\nu/C^2 $, 
it follows that $\Lambda_f =  \Lambda_g/C^2$,
which gives an algebraic equation for $C$, 
\begin{eqnarray}
&&
\cos^2\eta
\left(b_0 +3 b_1\,C+3 b_2\,C^2
+ b_3\,C^3\right)
\notag
\\
&&
={\sin^2\eta\over C}
\left(b_1 +3 b_2 C\,+3 b_3 C^2\,
+ b_4 C^3\,\right)
\label{cosmc}
\,.
\end{eqnarray}
It follows  also that the matter sources should be such that  
\bea
{\cal T}^{[{\rm m}]\mu}_{~~~~~\nu} =
T^{[{\rm m}]\mu}_{~~~~~\nu}\,/C^2
\,.
\eea
As a result, the independent equations are \eqref{GRa}, 
which are the same as in 
GR. 
In the present cosmological model, choosing
 the gauge where $\alpha=1$, 
these equations are obtained from \eqref{C1},\eqref{e1},\eqref{e2},
\bea                            \label{GR_Lambda}
\left(e^{3\Omega}\,\dot{\Omega}\right)^2&=&
\sigma_{+}^2+\sigma_{-}^2
+\frac{1}{3}\left(\Lambda_g+\rho_g \right)
e^{6\Omega}\,,\notag \\
\left(e^{3\Omega}\,{\dot{\Omega}}\right)^{\centerdot}
&=&\frac{1}{2}\,e^{3\Omega}(\rho_g-P_g), \notag \\
e^{3\Omega}\,\dot{\beta}_\pm&=&\sigma_\pm
\,,
\eea
where $\sigma_\pm$ are integration constants.
Solutions of these equations 
are reviewed in the Appendix. 
The solution for the f-metric is 
\bea
{\cal A}=C\,,~~
e^{\cal W}=Ce^\Omega\,,~~
{\cal B}_\pm=\beta_\pm
\,.
\eea
As shown in the Appendix, if $\Lambda_g$ is positive then the solutions 
approach  the  de Sitter metric exponentially fast, such that at late times 
one has
\be
\Omega={H t}+O(e^{-3H t}),~~~~
\beta_{\pm}=
\beta_{\pm}(\infty)+O(e^{-3H t})\,,
\ee
with $H=\sqrt{\Lambda_g/3}$. 
It is worth noting that, since we are in the Bianchi type I, 
the asymptotic values of the anisotropies,
$\beta_{\pm}(\infty)$, 
can be set to zero via rescaling the spatial coordinates,
which would not be true for other Bianchi types. 

If we choose the coupling constants $b_k$ 
according to 
Eq.~(\ref{bbb}), then Eq.\eqref{cosmc} factorizes as 
\be                     \label{PP}
(C-1)P_3(C)=0,
\ee 
where $P_3(C)$ is a cubic polynomial,
\bea
P_3(C)=(c_3+c_4)C^3+(3-5c_3+(\chi-2)c_4)C^2~~~~~~~  \\
+((4-3\chi)c_3+(1-2\chi)c_4-6)C+(3c_3+c_4-1)\chi,~~\notag 
\eea
with $\chi=\tan^2\eta$, 
while 
\bea
\Lambda_g&=&\cos^2\eta(1-C)  \\
&\times&((c_3+c_4)C^2+(3-5c_3-2c_4)C+4c_3+c_4-6). \notag
\eea
Depending on values of $c_3,c_4,\eta$, 
the equation \eqref{PP} can have up to four real roots. 
For example, for $c_3=1$, $c_4=0.3$, $\eta =1$ the four roots are
\be                            \label{Cval}
C=\{-2.245;\,1;\,0.068;\,3.616\},~~
\ee
and the corresponding 
\be                     \label{Lval}
{\Lambda_g}=\{10.126;\,0;\,-0.509;\,-4.505\}.
\ee  
As a result, we have four different solutions with four different
values of the cosmological constant, which can be positive, negative, or zero. 
Choosing $C=-2.245$ gives the de Sitter solution with the Hubble rate 
\be                         \label{Hubble}
H=\sqrt{\Lambda_g/3}=1.837.
\ee 
Below we shall find this value again 
for more complex solutions. 
If we choose $C=1$, 
then $\Lambda_g=\Lambda_f=0$ and the 
two metrics and  matter sources are identical:
\be
g_{\mu\nu}=f_{\mu\nu}\,,~~~~
\rho_g=\rho_f\,,~~~~P_g=P_f.
\ee
In vacuum, $\rho_g=P_g=0$, 
the solution is either Minkowski metric if $\sigma_\pm=0$
or the Kasner metric for non-zero $\sigma_{\pm}$. 

\subsection{Solutions with equal anisotropies}
If the two metrics are different
and the matter sources are not adjusted to be the same,
then the simplest solutions of equations \eqref{C1}--\eqref{e4}
are of the FLRW type \cite{Volkov:2011an}, in which case anisotropies vanish, 
\be
\beta_\pm={\cal B}_\pm=0.
\ee 
It turns out that one can obtain also more general 
solutions for which the anisotropies do not vanish but are the same  
 in both sectors, 
\be                                 \label{anis}
\beta_{\pm}={\cal B}_{\pm}\ne 0.
\ee
It is instructive to describe at the same time both 
isotropic solutions and solutions with equal anisotropies. 

The key point is that configurations with equal anisotropies extremize 
the potentials
$U_g$ and ${\cal U}_f$ so that their derivatives 
in \eqref{e2},\eqref{e4} vanish, 
\bea
&&
{\partial U_g\over \partial \beta_\pm}\Big{|}_{\beta_\pm={\cal B}_\pm}=0\,,
{\partial {\cal U}_f\over \partial \beta_\pm}\Big{|}_{\beta_\pm={\cal B}_\pm}
=0\,,
\nonumber \\
&&
{\partial U_g\over \partial {\cal B}_\pm}\Big{|}_{\beta_\pm={\cal B}_\pm}=0\,,
{\partial {\cal U}_f\over \partial {\cal B}_\pm}
\Big{|}_{\beta_\pm={\cal B}_\pm}=0\,.                       \label{potmin}
\eea
As a result, Eqs.\eqref{e2},\eqref{e4} can be integrated to give 
\be                       \label{bet}
e^{3\Omega}\,\frac{\dot{\beta}_{\pm}}{\alpha}=\sigma_\pm,~~~~
e^{3\cal W}\,\frac{\dot{\cal B}_{\pm}}{\cal A}=S_\pm,~~~~
\ee
with constant $\sigma_\pm$, $S_\pm$. A particular choice 
\be
\sigma_{\pm}=0,~~~~{\cal S}_{\pm}=0,
\ee
corresponds to isotropic FLRW cosmologies, since Eq.\eqref{bet} requires 
 in this case
that  
$\beta_{\pm}={\cal B}_{\pm}$ are constants, which can be set to zero 
by rescaling the spatial coordinates. 

If $\sigma_\pm,S_\pm$ do not vanish then,  
since $\dot{\beta}_\pm=\dot{\cal B}_\pm$, 
taking the ratio of the two expressions in \eqref{bet} gives the relation 
\be                                  \label{rat}
e^{3({\cal W}-\Omega)}\,\frac{\alpha}{\cal A}=
\frac{{\cal S}_{+}}{\sigma_{+}}=\frac{{\cal S}_{-}}{\sigma_{-}}\equiv C^2.
\ee
As we shall see, the constant here should be positive,
so that it is denoted by $C^2$. 

In view of \eqref{potmin}, the constraint \eqref{C:a} reduces to 
\be                              \label{Ca}
{\cal A}\,\dot{\Omega}\frac{\partial {\cal U}_f}{\partial\Omega}=
\alpha\,\dot{{\cal W}}\frac{\partial U_g}{\partial{\cal W}}\,,
\ee
which can be transformed to 
\be                         \label{factors}
\left[\alpha\left(e^{\cal W}\right)^
\cdot-{\cal A}\left(e^\Omega\right)^\cdot\right]
\left(b_1 +2b_2e^{{\cal W}-\Omega}+
b_3 e^{2({\cal W}-\Omega)}\right)=0. 
\ee
Depending on which factor here vanishes, there are two 
solution branches, which we shall call generic and special. 

\subsubsection{\bf Generic solutions}
Let us first consider the case where the first 
factor in \eqref{factors} vanishes, 
\be                           \label{first} 
\alpha\left(e^{\cal W}\right)^
\cdot-{\cal A}\left(e^\Omega\right)^\cdot=0. 
\ee 
Denoting 
$$
\xi=e^{{\cal W}-\Omega},
$$
one has 
\be                           \label{Wdot0}
\dot{\cal W}=\frac{\cal A}{\xi \alpha}\,\dot{\Omega}\,.
\ee
The remaining equations to be solved are the constraints \eqref{C1},\eqref{C2},
which reduce to 
\bea
\dot{\Omega}^2&=&\alpha^2\left\{\sigma^2 e^{-6\Omega}
+\frac{\Lambda_g(\xi)+\rho_g}{3}\right\}
\,,\label{C1a}  \\
\dot{\cal W}^2&=&{\cal A}^2\left\{{\cal S}^2 e^{-6\cal W}
+\frac{\Lambda_f(\xi)+\rho_f}{3}\,\right\}
\,.~~~~\label{C2a} 
\eea
Here $\sigma^2=\sigma_{+}^2+\sigma_{-}^2$ and ${\cal S}^2=
{\cal S}_{+}^2+{\cal S}_{-}^2$, while $\Lambda_g,\Lambda_f$
are obtained by replacing in Eq.\eqref{Lmbd} $C\to \xi$.

If we multiply \eqref{C2a} by $(\xi\alpha/{\cal A})^2$ 
and  then subtract from \eqref{C1a}
then, in view 
of \eqref{Wdot0}, the result will be 
\be                              \label{base0}
\xi^2\Lambda_f(\xi)-\Lambda_g(\xi)
=
\rho_g-\xi^2\rho_f+{\cal E}
\ee 
with 
\be                             \label{calF}
{\cal E}=3 \, e^{-6\Omega}\sigma^2\left[1-\left({C\over \xi}\right)^{4}
\right]
\,.
\ee

\noindent
{\sl (i) Isotropic solutions.--}
If $\sigma^2={\cal S}^2=0$ then 
${\cal E}=0$. 
Since  the matter densities determined by \eqref{mat1}
are $\rho_g(\Omega)$ and $\rho_f({\cal W})$ with
${\cal W}$ being function of $\Omega$ and $\xi$, 
Eq. \eqref{base0} gives 
an algebraic relation between 
$\xi$ and $\Omega$, which can be resolved to 
determine $\xi(\Omega)$.
Inserting  this function
into \eqref{C1a} gives the equation for $\Omega$,
\be                           \label{effect}
\dot{\Omega}^2=\frac{\Lambda_g(\xi(\Omega))+\rho_g(\Omega)}{3}
\equiv E^2_{\rm eff}(\Omega)\,,
\ee
where we have set $\alpha=1$.
Depending on the choice of the solution $\xi(\Omega)$, 
the form of the potential $E_{\rm eff}(\Omega)$ is different, 
leading either to self-acceleration or to other 
types of behavior \cite{Volkov:2011an}. 
 \begin{figure}[th]
\hbox to \linewidth{ \hss
	


\hspace{1 cm}	\resizebox{8cm}{5.2cm}{\includegraphics{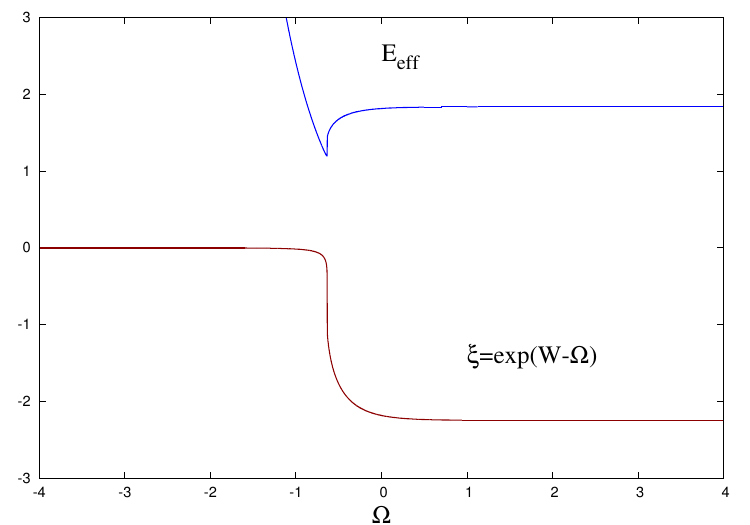}}

\hspace{5mm}

\hspace{1mm}
\hss}
\caption{{\protect\small The solution $\xi(\Omega)$ of the 
algebraic equation \eqref{base0}  
for $c_3=1$, $c_4=0.3$, $\eta=1$,
$\rho_g=0.25\times e^{-4\Omega}+0.25\times e^{-3\Omega}$,
$\rho_f=0$, and the corresponding potential $E_{\rm eff}(\Omega)$ 
in \eqref{effect}.
 }}%
\label{Fig1}
\end{figure}
If one requires 
the graviton mass contribution to the total energy density, $\Lambda_g$,  
to be dominant at late times (accelerating phase) but small at early times
(matter dominated phase), then one has to have  \cite{Volkov:2011an}
\be
\xi(\Omega)={e^{\cal W}/e^\Omega} 
\to 0 ~~{\rm as}~~~e^\Omega\to 0\,,
\ee
as well as 
\be                              \label{b1}
b_1<0 \,.
\ee
The function $\xi(\Omega)$ is then always negative%
\footnote{$\xi$ can be negative, since 
$e^{\cal W}$ or $e^\Omega$ need not to be positive definite,
because only their squares appear in the metric coefficients. }.

An example of a self-accelerating solution is shown in Fig.\ref{Fig1}.
For $\Omega\to\infty $ one has 
\be                                                \label{example}
\xi(\Omega)\to -2.245,~~~~~~E_{\rm eff}(\Omega)\to H=1.837\,,
\ee
where 
$H$ is the Hubble expansion rate.
In Section \ref{general_Bianchi} we shall study more general 
solutions with exactly the same late-time behavior.

\noindent
{\sl (ii) Anisotropic solutions.--}
If  ${\cal S}^2=\sigma^2C^2\neq 0$, 
then combining \eqref{first} with \eqref{rat} gives 
\be
\dot{\Omega}e^{-2\Omega}=C^2\dot{\cal W}e^{-2\Omega}\,,
\ee
from which, denoting the 
integration constant by $\nu$, 
we find 
\be                                 \label{xiK}
\xi=\pm \frac{C}{\sqrt{1-\nu e^{2\Omega}} }\,.
\ee
If $\nu$ does not vanish, then  
$$
{\cal E}=3\nu \sigma^2 e^{-4\Omega}\left(2-
\nu e^{2\Omega}\right)
\,. 
$$
Inserting this to \eqref{base0} gives an algebraic relation
between $\xi$ and $\Omega$, which, in view  of \eqref{xiK},
becomes a relation between $\Omega$ and $C$. 
As a result, $\Omega$ should be constant 
and no dynamical solutions exist in this case.

If $\nu=0$ then $\xi=C$ and Eq.\eqref{rat}
gives  ${\cal A}=C\alpha$
so that the two metrics are proportional, 
\be                         
f_{\mu\nu}=C^2g_{\mu\nu},
\ee
and ${\cal E}=0$.  
This case was discussed above in 
Section \ref{GR}, where we saw that the matter sources should be 
fine-tuned,
$\rho_g=C^2\rho_f$.

\subsubsection
{\bf Special solutions}
Let us now consider the case where the second
factor in \eqref{factors} vanishes,
\be                            \label{eqb}
b_3 \xi^2+2b_2 \xi+b_1=0,
\ee 
implying that $\xi$ is constant.
Since it should be real,
one has to have $b_2^2-b_1 b_3\geq 0$. 
The equations to be solved are again \eqref{C1a} and \eqref{C2a},
where now $\dot{\cal W}=\dot{\Omega}$. 
One has either  $\sigma^2={\cal S}^2=0$ in the isotropic case or,  
in view of \eqref{rat},
\be                      \label{ref1}
{\cal A}=\frac{\xi^3}{C^2}\,\alpha
\,,
\ee
in the anisotropic case. 
 In both cases, 
taking the difference of \eqref{C1a} and \eqref{C2a}
gives 
\be                           \label{calA}
{\cal A}=\sqrt{\frac{\Lambda_g(\xi)+\rho_g }{\Lambda_f(\xi)+\rho_f}}\,\alpha
\,.
\ee
In the isotropic case this completes the procedure -- solution for $g_{\mu\nu}$ 
is the same as in General Relativity 
with the cosmological constant $\Lambda_g(\xi)$ and matter $\rho_g$.
The second metric is obtained using the values of $e^{\cal W}=\xi e^\Omega$
and of  ${\cal A}$
from \eqref{eqb},\eqref{calA}. 
  The solution exists  if only
the argument of the square root in \eqref{calA} is positive, which condition
turns out to be rather restrictive. For example, 
if the parameters $b_k$ are chosen according 
\eqref{bbb}, then one finds 
\be
\frac{\Lambda_g(\xi)}{\cos^2\eta}
+\xi^2\frac{\Lambda_f(\xi)}{\sin^2\eta}=-(1-\xi)^2\,,
\ee
so that if $\Lambda_g>0$ then $\Lambda_f<0$, therefore only solutions
with $\rho_f>-\Lambda_f$ are allowed.  

For anisotropic solutions Eqs.\eqref{ref1} and \eqref{calA} give 
\be                           \label{calAaa}
C^2=\xi^3\sqrt{\frac{\Lambda_f(\xi)+\rho_f }{\Lambda_g(\xi)+\rho_g}}\,.
\ee
The argument of the square root here should be positive and, in addition, 
constant. Since, by our assumption, $\rho_g,\rho_f$ do not contain constant
contributions, they should be proportional, $\rho_f=(\Lambda_f/\Lambda_g)\rho_g$. 
This excludes self-accelerating solutions with $\Lambda_g>0$, 
since if $\Lambda_g$ and $\rho_g$
are positive, then 
both $\Lambda_f$ and $\rho_f$ are negative and the argument of the 
square root is negative.

\section{More general Bianchi types}
\label{general_Bianchi}
\setcounter{subsubsection}{0}
For the general Bianchi class A models we choose 
\bea
ds_g^2=&-&\alpha(t)^2dt^2+ \sum_{a=1,2,3}\alpha^2_a(t)\,\omega^a\otimes\omega^a\,,\notag \\
ds_f^2=&-&{\cal A}(t)^2dt^2+ \sum_{a=1,2,3}{\cal A}^2_a(t)\,\omega^a\otimes \omega^a\,,
\eea
where the vectors dual to $\omega^a$ do not commute, so that the 
spatial curvature do not vanish.  
Few solutions then can  be found in a closed form.

\subsubsection{\bf Recovering General Relativity}

Assuming again the two metrics and their sources
to be proportional, $f_{\mu\nu}=C^2 g_{\mu\nu}$, 
we obtain the GR equations,
\bea                            \label{GR1a}
\dot{\Omega}^2&=&\dot{\beta}_{+}^2+\dot{\beta}_{-}^2
-\frac16\,\R+\frac{\Lambda_g+\rho}{3}\,,  \\
\left(e^{3\Omega}\dot{\beta}_\pm\right)^\cdot&=&
\frac{e^{3\Omega}}{12}\frac{\partial \R}{\partial \beta_\pm}\,, \label{GR1b}
\eea
with $\R$ given by \eqref{R} and $\Lambda_g,C$ defined by \eqref{Lmbd},\eqref{cosmc}.
One can choose $C=1$, in which case $\Lambda_g=0$.

One cannot integrate the equations for $\beta_\pm$ analytically, 
unless one assumes that $\beta_\pm=0$, which is however consistent
 with \eqref{GR1b}
only when $n^{(1)}=n^{(2)}=n^{(3)}\equiv k$, which corresponds to
 the Bianchi I,IX for $k=0,1$, respectively.  
Eq.\eqref{GR1a} then becomes 
\be                            \label{GR1aa}
\dot{\Omega}^2=
-\frac{k}{4}\,e^{-2\Omega}+\frac{\Lambda_g+\rho}{3},
\ee
which describes the spatially flat $(k=0)$ or spatially closed $(k=1)$ 
FLRW universe.

\subsubsection{\bf The isotropic case}

If the the anisotropies vanish but the 
metrics and their sources are not proportional, then 
instead of \eqref{GR1aa} one obtains  
\be                           \label{effective}
\dot{\Omega}^2=-\frac{k}{4}\,e^{-2\Omega}+E^2_{\rm eff}(\Omega)\,,
\ee
where $E_{\rm eff}(\Omega)$ is the same as in \eqref{effect}. 
It is instructive to rewrite this equation as
\be                         \label{VVa}
\dot{\bf a}^2+V({\bf a})=-k
\ee
with ${\bf a}=2e^\Omega$ and $V=-{\bf a}^2E^2_{\rm eff}$. 
Since $\dot{\bf a}^2>0$, 
the solution is restricted to the regions where 
\be                         \label{VVb}
V({\bf a})\leq -k\,.
\ee
 \begin{figure}[th]
\hbox to \linewidth{ \hss
	

\hspace{1 cm}	\resizebox{8cm}{5.2cm}{\includegraphics{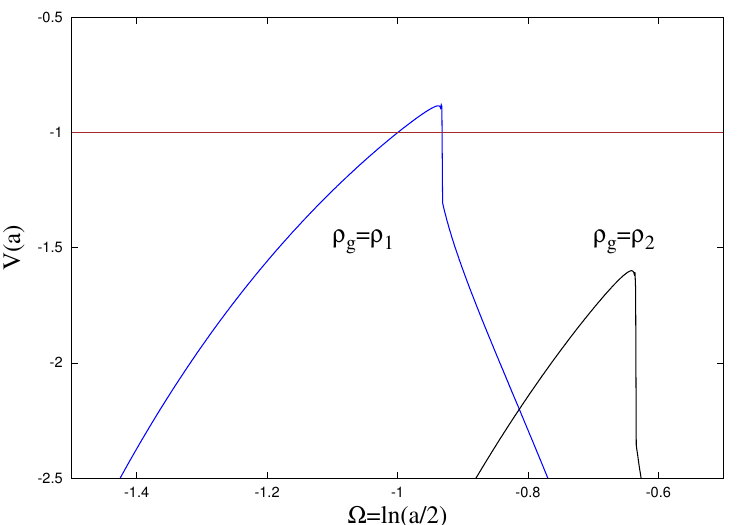}}

\hspace{5mm}

\hspace{1mm}
\hss}
\caption{{\protect\small Potential $V({\bf a})$ in \eqref{VVa}
for isotropic solutions for different choices 
of the matter density: $\rho_1=0.1\times e^{-4\Omega}$ and 
$\rho_2=0.25\times e^{-4\Omega}+0.25\times e^{-3\Omega}$ and
$\rho_f=0$. In the Bianchi IX case the motion is restricted
to the region $V<-1$, so that for $\rho_g=\rho_1$ one has a recollapsing
solution corresponding to the region on the left from the barrier,
and a bounce solution corresponding to the region on the right. 
 }}%
\label{Fig2}
\end{figure}
To construct $V({\bf a})$ one should resolve the algebraic equation 
\eqref{base0} (with ${\cal E}=0$) to obtain $\xi(\Omega)$ 
for chosen $\rho_g,\rho_f$,
 and then insert the result to \eqref{effect}
to obtain $E_{\rm eff}(\Omega)$. 
In Fig.\ref{Fig2} the result  is shown for two different choices of $\rho_g$
and for $\rho_f=0$. One can see that, since $V<0$, 
for $k=0$  (Bianchi I) the motion covers the 
whole region ${\bf a}\geq  0$, including the initial
 singularity at ${\bf a}=0$. The universe therefore 
expands forever starting from zero size. 

The same is true for $k=1$ (Bianchi IX), 
but if only $\rho_{g}$ is large enough ($\rho_g=\rho_2$),
so that the total amount of matter in the universe is sufficient.  
If $\rho_{g}$ is small ($\rho_g=\rho_1$) then the total `energy' is not enough
to pass over the barrier, and the motion is divided  
 into two regions. 
The region on the left from the barrier corresponds to universes 
 which start at the singularity,
then expand up to a maximal size, reflect from the barrier, and recollapse. 
The region on the right from the barrier describe universes that shrink 
from infinity down to a minimal non-zero size, then bounce from the barrier
 and expand again to infinity\footnote{
The similar bounce is found in de Sitter solution for the closed 
universe, i.e., $ {\bf a}\propto \cosh (Ht)$.}. 
As we shall see later,
 these features apply also when the anisotropies
 are taken into account.

\subsection{Generic solutions -- integration procedure} 

As we have seen, exact solutions can be found either 
in the isotropic case of the Bianchi I and IX, or when 
the anisotropies are equal, uniquely for the Bianchi I.  
In all other cases we are bound to use the numerical analysis. 
Our proceure will be described below, while its outcome can be 
summarized already now: it turns out that the solutions with equal 
anisotropies in the two sectors play the role of
asymptotic states to which the spacetime approaches at late times.

In order to integrate the equations numerically, we first 
of all convert the 
 equations to the form of a  dynamical system. 
Let us introduce the variables  
\bea                          \label{vars}
y_0&=&e^\Omega,~~~~~~~y_1=e^{\beta_{+}},~~~~~~~y_2=e^{\sqrt{3}\beta_{-}}, \\
y_3&=&e^{\cal W},~~~~~~~y_4=e^{{\cal B}_{+}},~~~~~~~y_5=e^{\sqrt{3}{\cal B}_{-}},\notag \\
y_6&=&\frac{e^{3\Omega}}{\alpha}\,\dot{\Omega},~~~~
y_7=\frac{e^{3\Omega}}{\alpha}\,\dot{\beta}_{+},~~~
y_8=\frac{e^{3\Omega}}{\alpha}\,\dot{\beta}_{-},~  \notag \\
y_9&=&\frac{e^{3{\cal W}}}{\cal A}\,\dot{{\cal W}},~~~~
y_{10}=\frac{e^{3{\cal W}}}{\cal A}\,\dot{{\cal B}}_{+},~~~
y_{11}=\frac{e^{3{\cal W}}}{\cal A}\,\dot{{\cal B}}_{-}.\notag 
\eea
The second order field equations \eqref{e1}, \eqref{e2}, \eqref{e3}, \eqref{e4} 
can be represented in the first order form 
\be                         \label{dyn}
\dot{y}_N=F_N(\alpha,{\cal A},y_M)\,,
\ee
where $N,M=0,\ldots,11$ and 
$F_N(\alpha,{\cal A},y_M)$ are defined in the Appendix. 

The first order equations 
\eqref{C1}, \eqref{C2}, \eqref{C:a} are
constraints 
\be
{\cal C}_1(y_N)=0,~~~
{\cal C}_2(y_N)=0,~~~
{\cal C}_3(y_N)=0,~~~\label{CCONS}
\ee
with 
\begin{eqnarray*}                            \label{CCC}
{\cal C}_{1}&=& -y_6^2+y_7^2+y_8^2+2c^2y_0^3U_g
-\frac16\,y_0^6\,\R
+\frac13 y_0^6\,\rho_g
,~~~\notag \\
{\cal C}_{2}&=& -y_9^2+y_{10}^2+y_{11}^2+2s^2y_3^3{\cal U}_f
-\frac16\,y_3^6\,\Rcal
+\frac13 y_3^6\,\rho_f,\notag  \\
{\cal C}_3&=&
y_3^3\left(
y_6 y_0\frac{\partial}{\partial y_0}+
y_7 y_1\frac{\partial }{\partial y_1}+
\sqrt{3}y_8 y_2\frac{\partial }{\partial y_2}
\right){\cal U}_f  \\
&-&
y_0^3\left(
y_9 y_3\frac{\partial }{\partial y_3}+
y_{10} y_4\frac{\partial }{\partial y_4}+
\sqrt{3}y_{11} y_5\frac{\partial }{\partial y_5}
\right)U_g,\notag 
\end{eqnarray*}
where $U_g$, ${\cal U}_f$, $\R$, $\Rcal$, $\rho_g$, $\rho_f$
 are defined  in the Appendix.

In order to implement the constraints and also determine the 
lapses $\alpha$ and ${\cal A}$, we remember 
that the third constraint in \eqref{CCONS} 
is obtained as the condition
of propagation of the first two. Specifically, calculating the 
derivatives
\be                            \label{CC1a}
\dot{\cal C}_1=\sum_{N=0}^{11}\frac{\partial {\cal C}_1}{\partial y_N}\,F_N
\,,~~~~~
\dot{\cal C}_2=\sum_{N=0}^{11}\frac{\partial {\cal C}_2}{\partial y_N}\,F_N\,
\ee 
gives expressions proportional to
 ${\cal C}_3$. 
Therefore, if the third constraint is fulfilled, ${\cal C}_3=0$, then 
$\dot{\cal C}_1=\dot{\cal C}_2=0$,
so that it is sufficient to require ${\cal C}_1={\cal C}_2=0$
only at the initial moment of time. 

Now, what guarantees that the third constraint propagates itself ?
In fact, calculating its time derivative 
\be                            \label{CC1b}
\dot{\cal C}_3=\sum_{N=0}^{11}\frac{\partial {\cal C}_3}{\partial y_N}\,F_N\,,
\ee
and setting the result
 to zero 
we discover a non-trivial relation of the form 
\be                       \label{QQQ}
\alpha X_\alpha +{\cal A} X_{\cal A}=0,
\ee
where $X_\alpha$ and $X_{\cal A}$ are rather complicated functions of $y_N$
(we do not show them explicitly). 
This is the condition of propagation of the third constraint, 
which in turn guarantees that the first two constraints propagate. 
This condition determines the lapse ${\cal A}$,
\be                                 \label{AAA}
{\cal A}=-\frac{X_\alpha}{X_{\cal A}}\,\alpha\,. 
\ee

Our procedure is then as follows: we integrate the 12 
equations \eqref{dyn} with ${\cal A}$ determined
by \eqref{AAA}, where we can use the gauge condition
\be
\alpha=1.
\ee
The three constraints are imposed 
 at the initial time moment,
and the above consideration guarantee that they
are fulfilled for all times.

To determine the initial data, we choose 9 out of 12 amplitudes $y_N$
and give them some initial values. It is convenient to choose
\be
y_0,~~y_1,~~y_2,~~y_4,~~y_5,~~y_7,~~y_8,~~y_{10},~~y_{11}
\ee
which determine $\Omega$, $\beta_{\pm}$, ${\cal B}_{\pm}$,
$\dot{\beta}_\pm$, $\dot{B}_{\pm}$ at $t=0$. Having chosen these 9 values,
the values of $y_3,y_6,y_9$ determining ${\cal W},\dot{\Omega},\dot{W}$ 
are obtained by numerically solving the three constraint equations
 \eqref{CCONS}. 
Having done these, all initial data are determined, and there remains 
just to integrate forward the 12 equations \eqref{dyn} using
the numerical routines of \cite{Press:2007:NRE:1403886}.

To check our procedure, we specialize to the Bianchi I type by setting in the 
equations $n^{(1)}=n^{(2)}=n^{(3)}=0$ and also 
switch off the matter, $\rho_{g}=\rho_{f}=0$.
We then first 
consider the case where all anisotropies and their time derivatives
initially vanish, 
$$
y_1=y_2=y_4=y_5=1\,,~~
y_7=y_8=y_{10}=y_{11}=0.
$$  
The numerical integration then reproduces the de Sitter solution,
for which these conditions are preserved for all times.
Next, if we initially set 
$$
y_1=y_4\neq 1,~~~y_2=y_5\neq 1,~~~
y_7=y_8=y_{10}=y_{11}=0,
$$  
then these conditions are also preserved  
for all times. This corresponds to the solution with 
$\sigma_\pm={\cal S}_\pm=0$
but with constant values of 
$\beta_{+}={\cal B}_{+}$ and $\beta_{-}={\cal B}_{-}$. 
This is again the de Sitter solution, since constant anisotropies
can be removed by rescaling the spatial coordinates.

As the next step, we initially choose 
\bea
y_1&=&y_4\neq 1,~~~y_2=y_5\neq 1,~~~
y_7=\sigma_{+},~~y_8=\sigma_{-},\notag \\
y_{10}&=&C\sigma_{+},~~~y_{11}=C\sigma_{-},~~~~C={y_3^3}/{({\cal A}y_0^3)}\,.
\eea 
The numerical solution then again preserves these conditions at all times, 
which corresponds to the analytic solutions with equal anisotropies. 

\subsection{Numerical results}

After consistency checks, we turn to 
the generic case.  
We switch on the matter by setting in the equations 
$\rho_{g}\neq 0$ and $\rho_{f}\neq 0$.
We choose an initial value for $y_0=e^{\Omega}$ 
and some values for
$y_1=e^{\beta_{+}}$, 
$y_2=e^{\sqrt{3}\beta_{-}}$,
$y_3=e^{{\cal B}_{+}}$, 
$y_4=e^{\sqrt{3}{\cal B}_{-}}$
assuming that $\beta_\pm$ and ${\cal B}_{\pm}$ are not too large. 
In addition, we choose small enough values for  
$y_7,y_8,y_{10},y_{11}$ 
determining $\dot{\beta}_\pm$, $\dot{\cal B}_\pm$,  
so that the initial configuration is not too anisotropic.

We consider all Bianchi class A types.
The values of the parameters $n^{(a)}$ are given in Table I. 
\begin{table}[h]
\begin{tabular}{|c|c|c|c|c|c|c|}
\hline
    & ~I~~ & ~II~ & ~VI$_0$ & VII$_0$ & VIII & ~IX~ \\
\hline
$n^{(1)}$ & $0$ & $1$ &  $1$ & $1$ & $1$ & $1$ \\
$n^{(2)}$ & $0$ & $0$ & $-1$ & $1$ & $1$ & $1$ \\
$n^{(3)}$ & $0$ & $0$ &  $0$ & $0$ & $-1$ & $1$ \\
\hline
\end{tabular}
\caption{Values of $n^{(a)}$ for Bianchi class A types}
\end{table}


We find that 
if the parameters are chosen such that the universe 
enters the self-accelerating regime, then generic solutions 
rapidly approach  the state with equal and constant anisotropies, 
$\beta_{+}={\cal B}_{+}$ and $\beta_{-}={\cal B}_{-}$. 
For Bianchi I type one can scale away these constant values by rescaling 
the spatial coordinates, but this cannot be done for other Bianchi types. 
As a result, the final state of the universe is anisotropic, even though 
the metric amplitudes $\dot{\Omega}$, $\dot{\cal W}$, ${\cal A}$  approach 
constant values, as for the isotropic universe.

\begin{figure}[h]
\hbox to \linewidth{ \hss
	

\hspace{1mm}
	\resizebox{8cm}{5.2cm}{\includegraphics{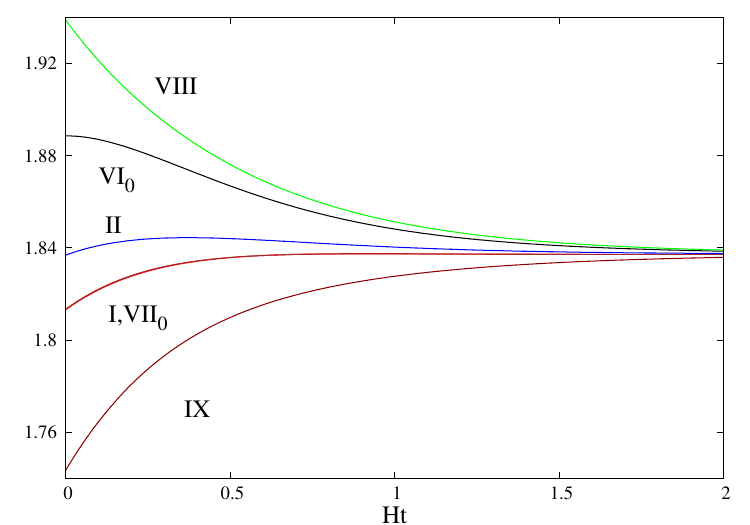}}

\hspace{1mm}
\hss}
\caption{{\protect\small The Hubble parameter $\dot{\Omega}$ for solution 
for all
Bianchi types of class A. 
 }}%
\label{Fig3}
\end{figure}

\begin{figure}[h]
\hbox to \linewidth{ \hss
	
	\resizebox{8cm}{5.2cm}{\includegraphics{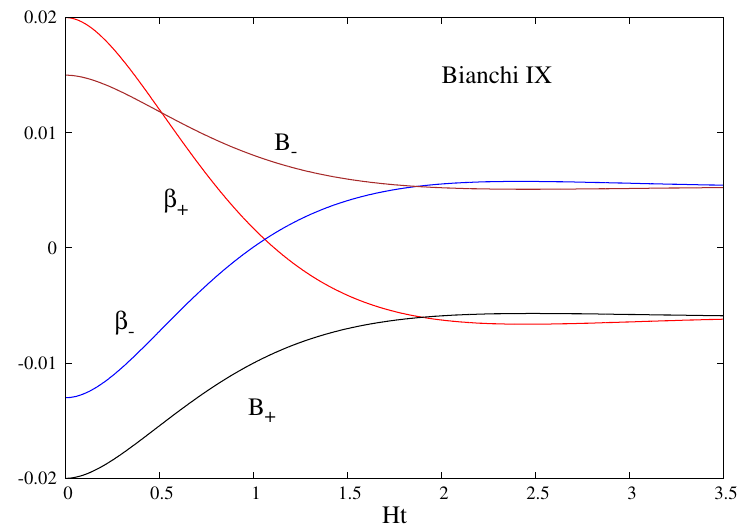}}
\hspace{1mm}

\hspace{1mm}
\hss}
\caption{{\protect\small Anisotropy amplitudes profiles 
for Bianchi IX. For other Bianchi types the picture is similar. 
 }}%
\label{Fig4}
\end{figure}

\begin{figure}[h]
\hbox to \linewidth{ \hss
	

\hspace{1mm}
	\resizebox{8cm}{5.2cm}{\includegraphics{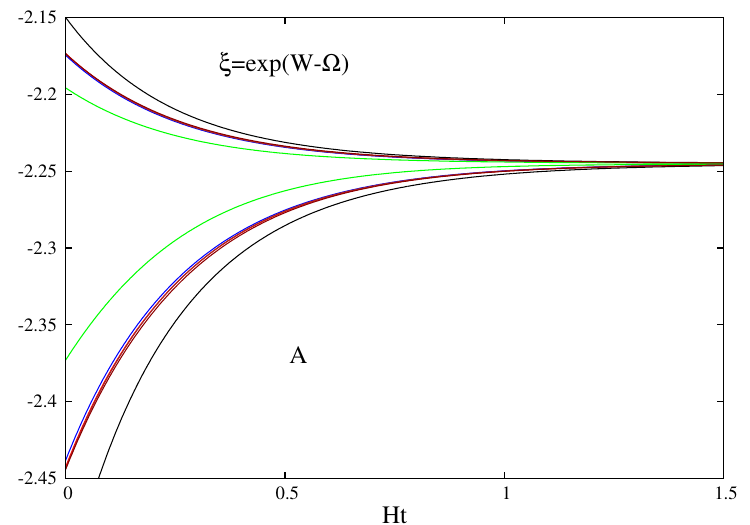}}
	
\hspace{1mm}
\hss}
\caption{{\protect\small The ratio of the two scale factors  
 $\xi=e^{{\cal W}-\Omega}$ and the lapse amplitude 
${\cal A}$ for all Bianchi types. 
 }}%
\label{Fig5}
\end{figure}

\begin{figure}[h]
\hbox to \linewidth{ \hss
	

\hspace{1mm}

	\resizebox{8cm}{5.2cm}{\includegraphics{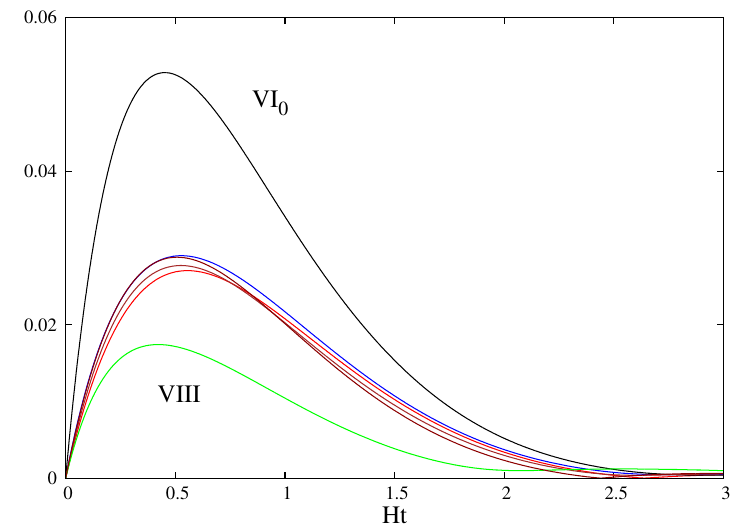}}
	
\hspace{1mm}
\hss}
\caption{{\protect\small The relative magnitude of shears $\Sigma=
(\dot{\beta}_{+}^2+\dot{\beta}_{-}^2)^{1/2}/\dot{\Omega}$
for all Bianchi types. 
 }}%
\label{Fig6}
\end{figure}

\begin{figure}[h]
\hbox to \linewidth{ \hss
	

\hspace{1mm}

	\resizebox{8cm}{5.2cm}{\includegraphics{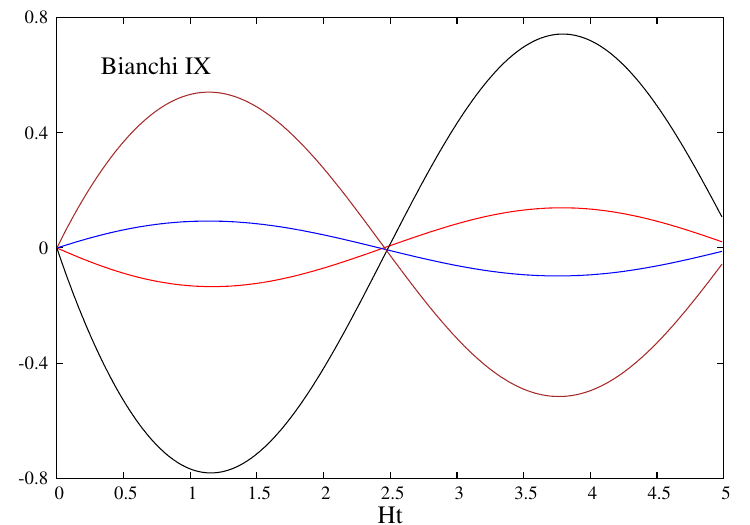}}
	
\hspace{1mm}
\hss}
\caption{{\protect\small The rescaled shears $e^{3\Omega/2}\dot{\beta}_\pm$ 
(smaller amplitudes) 
and $e^{3\Omega/2}\dot{\cal B}_\pm$ (larger amplitudes)
for the Bianchi IX solution, for other Bianchi types the picture
is similar. 
 }}%
\label{Fig7}
\end{figure}

Some examples of our solutions are presented 
in Figs.\ref{Fig3}--\ref{Fig9}
for the parameter values
$c_3=1$, $c_4=0.3$, $\eta=1$,
 which are the same as for the FLRW self-accelerating
solution in Fig.\ref{Fig1}. We assume the f-sector to be empty,
while the g-sector to contain a radiation and a non-relativistic 
matter described by  
$
\rho_g=0.25\times  e^{-4\Omega}+0.25\times e^{-3\Omega}\,,
$ 
which is the same function as in Figs.\ref{Fig1},\ref{Fig2}. 
There is nothing special about this choice, 
since for other parameter values solutions are qualitatively the same.
(We notice that $\rho_g\sim 1$ for $\Omega=0$, 
therefore, if $m\sim 10^{-33}{\rm eV}$, the dimensionful 
energy density in \eqref{Tm} is $\sim m^2/\kappa^2\sim 10^{-10} ({\rm eV})^4$,
which is about the present density of the universe).
The universe will accelerate if 
$c_3,c_4$ are chosen such that $b_1<0$ (see Eq.\eqref{b1}) while 
$\rho_g$ is large enough in order  that the system could travel over
the potential barrier as shown in Fig.\ref{Fig2}.  It is also worth noting
that the time scale  in Figs.\ref{Fig3}--\ref{Fig9} is the
Hubble time, $1/H$.

The initial value of the scale factor is $\Omega=0$ and 
the initial anisotropies are 
$\beta_{+}=0.02$,
$\beta_{-}=-0.013$, 
${\cal B}_{+}=-0.02$, 
${\cal B}_{-}=0.015$,
while the initial values of $\dot{\beta}_\pm$, $\dot{\cal B}_\pm$ are  
set to zero. 
Again, changing these values 
does not change the qualitative behavior of  the solutions. 
The values of ${\cal W}$, $\dot{\Omega}$, $\dot{\cal W}$
are determined by the constraints \eqref{CCONS}. 

As seen in Fig.\ref{Fig3}, for all Bianchi types 
the universe expansion rate 
rapidly approaches the constant value, $\dot{\Omega}\to H=1.837$
(the same value as in Eq.\eqref{Hubble}),
so that the universe approaches the de Sitter phase. 
If we return for a moment to
 dimensionful quantities, the Hubble rate becomes ${\bf H}=Hm$,
so that the cosmological 
constant is indeed due to the graviton mass.

The anisotropy amplitudes $\beta_\pm$, ${\cal B}_{\pm}$ 
approach 
constant values within an approximately twice Hubble time, $2H^{-1}$,
as is seen in Fig.\ref{Fig4}, where the solutions for the Bianchi IX are shown.   
The ratio of the two scale factors 
$\xi=e^{\cal W}/e^{\Omega}$ is shown in Fig.\ref{Fig5}. 
It is always negative for all solutions
and rapidly approaches the value $C=-2.245$, which is 
the same as in Eq.\eqref{Cval}. 
The lapse function ${\cal A}$, also shown in Fig.\ref{Fig6}, approaches the same value,
so that the two metrics become proportional at late times, $f_{\mu\nu}=C^2g_{\mu\nu}$,
as for the solutions 
 in section \ref{GR}.

Fig.\ref{Fig6} shows the relative magnitude of shears 
$$
\Sigma={(\dot{\beta}_{+}^2+\dot{\beta}_{-}^2)^{1/2}
\over \dot{\Omega}}
\,,
$$
which measures the relative contribution of the anisotropies to the total 
expansion rate. 
The maximal value of $\Sigma$ depends on the initial 
anisotropy values. However, setting the latter to zero
we obtain practically the same curves for the Bianchi types II,VI$_0$,VII$_0$,VIII,
since anisotropies cannot be zero in these cases, so that they are driven
to non-zero values by the cosmic expansion.  
The shears approach zero exponentially fast in Hubble units. 
However, if our universe entered the acceleration phase only recently, then one can expect 
that only 
few e-folding times have  elapsed since then, so that the shear energy 
is not necessarily small at present. 

The shears $\dot{\beta}_\pm$, $\dot{\cal B}_\pm$ 
multiplied by  $e^{3\Omega/2}$ 
oscillate with constant amplitudes, as shown in Fig.\ref{Fig7}, so that 
$\dot{\beta}_\pm\sim \dot{\cal B}_\pm\sim e^{-3\Omega/2}$.  
To understand this,  we remember that our configurations approach the solution  
with proportional metrics, $f_{\mu\nu}=C^2g_{\mu\nu}$, described in section \ref{GR},
so that the deviations from this solutions are small at late times. 
We therefore linearize the field equations with respect to the 
deviations, and solving the linearized equations we obtain
\be
\dot{\beta}_\pm\sim \dot{\cal B}_\pm\sim e^{-3Ht/2}\cos(H\omega t)
\ee
with 
\be
\omega^2=(b_1+2Cb_2+C^2b_3)\left(\frac{C\cos^2\eta}{H^2}+\frac{\sin^2\eta}{CH^2}\right)-\frac{9}{4}\,.
\ee
For our parameter values this gives $\omega=1.183$ and the oscillation
period in Hubble time units $T=2\pi/\omega=5.309$, in perfect
agreement with what is shown in Fig.\ref{Fig7}. 

Therefore, the shear contribution to the Hubble rate is 
\be
\dot{\beta}_{+}^2+\dot{\beta}_{-}^2\sim e^{-3\Omega}\sim 1/{\bf a}^3\,,
\ee
which falls off similarly to the energy density of a non-relativistic matter. 
In GR (see the Appendix) the shear falloff is much faster, 
$\dot{\beta}_{+}^2+\dot{\beta}_{-}^2\sim 1/{\bf a}^6$, corresponding
to a `stiff matter'. 
Since in the bigravity the shear contribution to the total
energy density increases slowly,  
this could have observational effects.

\subsubsection{Near singularity behavior}

\begin{figure}[h]
\hbox to \linewidth{ \hss
	


\hspace{1mm}
	\resizebox{8cm}{5.2cm}{\includegraphics{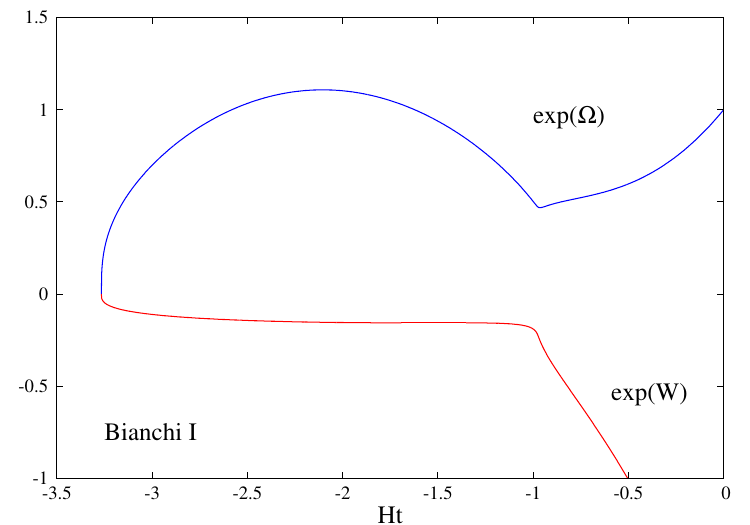}}
	
\hspace{1mm}
\hss}
\caption{{\protect\small The scale factors 
near the singularity. 
 }}%
\label{Fig8}
\end{figure}

It is interesting to 
see how the solutions continue to the past. 
Their parameters are chosen to avoid the bounce behavior, so that 
when continued to the negative $t$ region, they should 
hit a singularity at some point. 
The numerical simulations confirm these expectations and reveal that  
for all Bianchi types under consideration there is a singularity 
for $t<0$, where both $e^\Omega$ and $e^{\cal W}$ vanish. 
For the Bianchi I this is shown in Fig.\ref{Fig8}. 
Interestingly, 
in the negative $t$ region the solution shows a throat,
and $e^\Omega$ first expands before collapsing to zero. 

Let us consider in more detail the case 
of Bianchi IX.
In General Relativity Bianchi IX solutions reveal chaotic features. 
When approaching singularity, the metric coefficients $\alpha_a$ which measure 
the proper distances along the spatial axes (defined in \eqref{pm2}) 
show an infinite number of oscillations, whose 
 positions and amplitudes  are ergodic  \cite{PhysRevLett.22.1071,*Belinskii:1972}.  
Within an effective description, such a behavior is explained by a two dimensional
`billiard' motion of a particle reflecting from rigid walls.

  \begin{figure}[h]
\hbox to \linewidth{ \hss
	

\hspace{1mm}
	\resizebox{8cm}{5.2cm}{\includegraphics{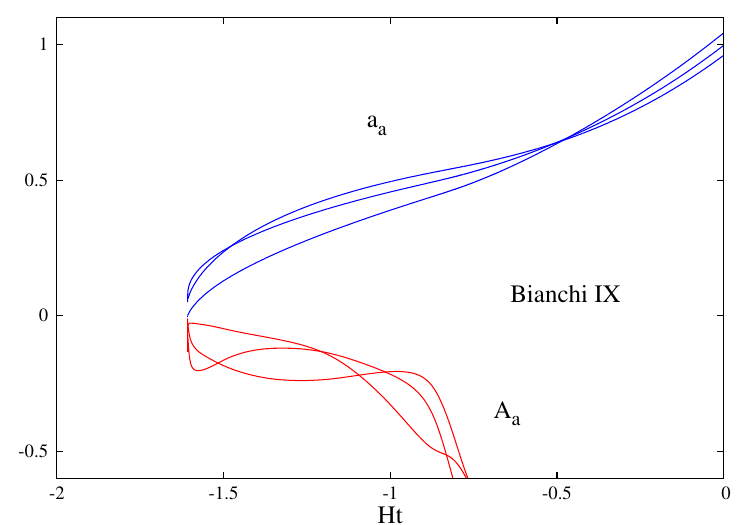}}
	
\hspace{1mm}
\hss}
\caption{{\protect\small The metric coefficients $\alpha_a$, ${\cal A}_a$
near the singularity. 
 }}%
\label{Fig9}
\end{figure}

At first glance, nothing similar is seen in our case. 
Fig.\ref{Fig9} shows the Bianchi IX solution continued to the past, 
and the singularity corresponds 
to a point where both $e^\Omega$ and $e^{\cal W}$ vanish, but in its 
vicinity coefficients $\alpha_a$, ${\cal A}_a$
defined by \eqref{pm2} approach zero without oscillations. 
However, zooming the picture one can see that oscillations actually start,
but not many of them are seen, 
since we cannot approach the singularity close enough
due to numerical errors. 

  \begin{figure}[h]
\hbox to \linewidth{ \hss
	


\hspace{1mm}
	\resizebox{8cm}{5.2cm}{\includegraphics{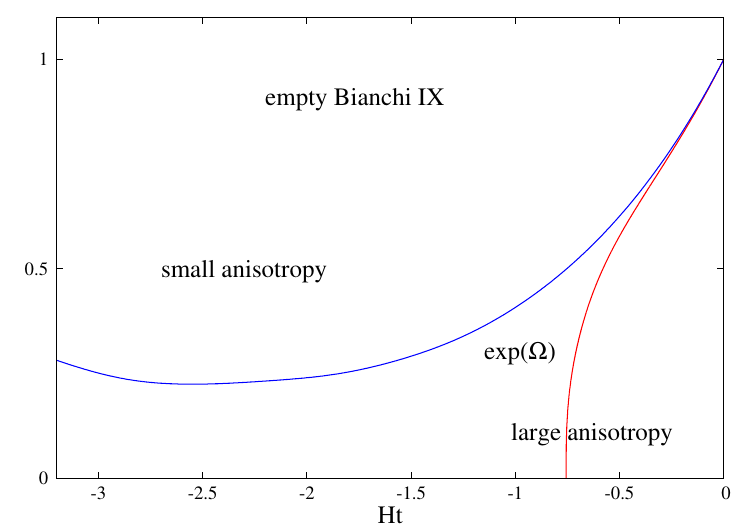}}
	
\hspace{1mm}
\hss}
\caption{{\protect\small 
Self-accelerating for $t>0$ Bianchi IX solutions continued 
to the $t<0$ region. Choosing at $t=0$  
$\beta_{\pm}\sim{\cal B}_{\pm}\sim 10^{-2}$, the solution is of regular
bounce type, but if 
$\beta_{\pm}\sim{\cal B}_{\pm}\sim 10^{-1}$ then the solution develops 
a singularity where $e^\Omega$ vanishes.
 }}%
\label{Fig10}
\end{figure}
The picture becomes more clear for solutions without matter.
Setting $\rho_g=\rho_f=0$, the simplest salf-accelerating solution 
is pure de Sitter, but it is of course non-singular. 
It can be obtained from Eq.\eqref{VVa}, which reduces to 
(remember that ${\bf a}=2e^\Omega$)
\be                                         \label{DS00}
\dot{\bf a}^2=H^2{\bf a}^2-1\,~~~~\Rightarrow~~~{\bf a}=\frac{1}{H}\,\cosh(t-t_0),
\ee
where $H=1.837 $ is the same as for all other solutions under considerations.  
Qualitatively, the singularity is avoided because 
${\bf a}$ cannot be too small, since otherwise the right hand side of the equation
would be negative. Therefore, ${\bf a}$ bounces back when it achieves the 
minimal value $1/H$. 

However, for non-zero anisotropy  the equation in \eqref{DS00} receives in the 
right hand side additional terms proportional to 
$\dot{\beta}_{+}^2+\dot{\beta}_{-}^2$, and these can keep the whole
expression on the right positive even if ${\bf a}\to 0$. Therefore, solutions with high
enough anisotropy can approach singularity. To verify this, we 
choose at $t=0$ two sets of initial values for the anisotropies, 
$\beta_{\pm}\sim{\cal B}_{\pm}\sim 10^{-2}$ and 
$\beta_{\pm}\sim{\cal B}_{\pm}\sim 10^{-1}$. 
When continued to the future, we see self-acceleration in both cases, 
but when continued to the past, we obtain the bounce behavior in the first 
case and a curvature singularity in the second case (see Fig.\ref{Fig10}). 
Therefore, the empty de Sitter spacetime 
becomes singular when too much anisotropy 
is added.

\begin{figure}[h]
\hbox to \linewidth{ \hss
	

%
\hspace{1mm}
	\resizebox{8cm}{5.2cm}{\includegraphics{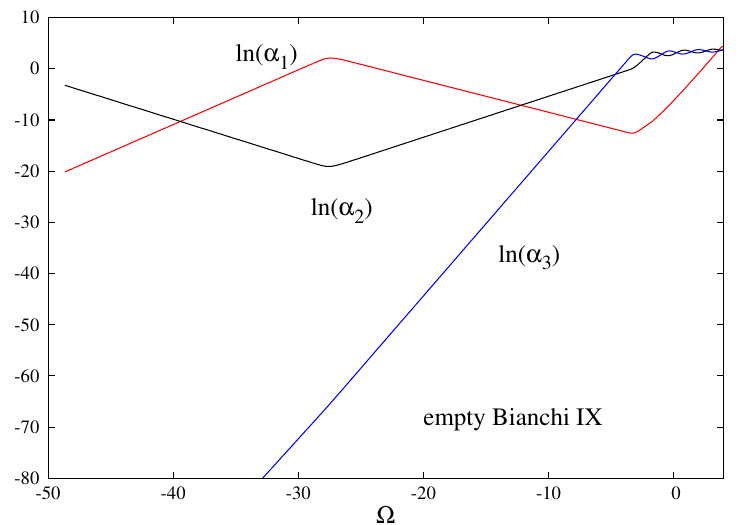}}
	
\hspace{1mm}
\hss}
\caption{{\protect\small The anisotropy parameters $\alpha_a$  
for the empty Bianchi IX near the singularity plotted against $\Omega$.
They show the typical 
`billiard' behavior characterized by a sequence of Kasner periods. 
For ${\cal A}_a$ the picture is similar. 
 }}%
\label{Fig11}
\end{figure}

It happens that for the empty Bianchi  IX solution we can  
approach the singularity 
much closer numerically, and in Fig.\ref{Fig11}
we show $\ln(\alpha_a)$ against $\Omega\propto \ln(t)$. 
This time one can clearly see the typical features of the billiard motion characterized by 
a sequence of Kasner-like periods. 
During each period one has $\alpha_a\propto t^{p_a}$, where $p_a$ fulfill Eq.\eqref{AKasner}. 
During the next period, 
the values ${p_a}$ change such that one of then remains positive, the 
corresponding amplitude continues to decrease toward singularity ($\alpha_3$ in the Fig.\ref{Fig10}),
while two other  $p_a$'s change sign, such that the increasing amplitude becomes
decreasing and vise versa.   
Of course, 
within our numerical approach we can capture only the beginning of the infinite sequence 
of Kasner-like steps, and so it would be interesting to develop an affective analytical
description.  If the mater is present, provided that $w<1$%
\footnote{The chaotic behavior in Bianchi IX disappears if a massless scalar field 
is included, since the effective equation of state in this case is $P=\rho$, so that 
near singularity
$\rho\propto e^{-6\Omega}$ 
is able to compete with 
the effect of anisotropies \cite{Belinskii:1973}. },
then its contribution 
to the equations is subleading as compared to that of the anisotropy terms, 
so that it cannot influence the chaotic behavior. Therefore, since the empty Bianchi IX
solutions are chaotic, so should be those with matter.    

\section{Concluding Remarks}
We have studied  anisotropic cosmologies in the ghost-free
bigravity assuming both metrics to be diagonal and of the 
same Bianchi type of class A. Including a source consisting of a radiation and 
a non-relativistic matter, we  considered generic initial data 
describing (not too large) anisotropic deformations of  
a FLRW  universe. We find
that the universe evolves into a state in which 
it expands with a constant Hubble rate proportional to 
the graviton mass, while the anisotropy parameters approach 
constant non-zero values. In the Bianchi I case
constant anisotropies can be scaled away be redefining the 
spatial coordinates, but not for other Bianchi types.  
For example, for the Bianchi IX solutions the constant $t$ spatial sections 
will be not round 3-spheres but squashed spheres.  

The conclusion is that generic self-accelerating cosmologies in bigravity 
are anisotropic. The anisotropy contribution to the total expansion rate
approaches zero exponentially fast in Hubble units, but if our universe entered 
the acceleration phase only recently, then one can expect that only 
few e-folding times have elapsed since then, so that the shear energy 
is not necessarily small. At late times the shear contribution to 
the total expansion rate  decreases much slower than in GR, 
only as an inverse cube of the size of the universe, 
which is the same  falloff rate as for a non-relativistic matter. 
Therefore, the anisotropy effect could be visible, 
although comparing with observations goes beyond the scope of this paper
(see \cite{Akrami:2013pna} for 
a data-fitting for the FLRW cosmologies with massive gravitons).

The fact  that the anisotropy contribution shows the same falloff rate 
as a cold dark matter suggests that the latter could in fact 
be the effect of the anisotropies.  If this were true, then it would follow that 
theories with massive gravitons could explain 
both the dark energy -- the cosmological term mimicked by the graviton mass, 
and also the cold dark matter mimicked by the anisotropies. 
However, it is unclear if this interpretation  can explain also 
the other properties of dark matter, as for example its clustering. 
In addition, the high isotropy of the cosmic microwave background implies that 
already at the recombination the metric anisotropies were small, therefore   
they are unlikely to explain 
almost a quarter of the total energy in the universe presently attributed to the 
dark matter. It would nevertheless be interesting to check if the 
anisotropy energy has a tendency for clustering, which could perhaps
be seen at the level of perturbations, but such an  analysis 
requires a separate study.

It would be interesting to develop an analytic description for the behavior 
of the Bianchi IX solutions near singularity. 
Yet
one more interesting open issue would be to see if the FLRW solutions with 
non-diagonal metrics studied in \cite{Koyama:2011xz,*Koyama:2011yg}, 
\cite{Chamseddine:2011bu}, \cite{D'Amico:2011jj}, \cite{Kobayashi:2012fz}, \cite{Volkov:2012cf,*Volkov:2012zb} could be
generalized for non-zero anisotropies.

\acknowledgments
We would like to thank John Barrow and Thibault Damour 
 for valuable comments.
KM was partially supported by the Grant-in-Aid for Scientific Research
Fund of the JSPS (C)  (No.22540291).
KM would like to thank the 
 University of Tours, the Yukawa Institute for Theoretical Physics, 
the University of Auckland
and the University of Canterbury 
where parts of this work were performed.

\appendix


\section{Bianchi I spacetimes 
in GR}
\renewcommand{\theequation}{{A}.\arabic{section}.\arabic{equation}}

In this Appendix, we discuss the Bianchi I spacetime in General Relativity 
for some matter types. 
Assuming the metric form
\bea
ds^2&=&- dt^2+\sum_{a=1,2,3}\alpha_a(t)^2
(dx^a)^2
\eea
where
$$
\alpha_1=e^{\Omega}e^{\beta_++\sqrt{3}\beta_-}
\,,~~
\alpha_2=e^{\Omega}e^{\beta_+-\sqrt{3}\beta_-}
\,,~~
\alpha_3=e^{\Omega}e^{-2\beta_+}\,,
$$
the Einstein equations 
reduce to 
\bea
\left(e^{3\Omega}\,{\dot{\Omega}}\right)^2&=&
\left(e^{3\Omega}\,{\dot{\beta}_{+}}\right)^2
+\left(e^{3\Omega}\,{\dot{\beta}_{-}}\right)^2
+\frac{1}{3}e^{6\Omega}\rho\,,\notag ~~~~\label{A1} \\
\left(e^{3\Omega}\,{\dot{\Omega}}\right)^{\centerdot}
&=&
\frac{1}{2}\,e^{3\Omega}(\rho-P),  \notag \label{A2}\\
\left(e^{3\Omega}\,{\dot{\beta}_{\pm}}\right)^{\centerdot}
&=&0,             \label{A3}
\eea
where $\rho$ and $P$ are the energy density and pressure, 
which fulfill 
the energy conservation condition,
\be
\dot{\rho}+3\dot{\Omega}\,(\rho+P)=0.
\ee
The last of Eqs. (\ref{A3}) can be integrated to give 
\bea
e^{3\Omega}\dot \beta_\pm= \sigma_\pm
\,,
\eea
where $\sigma_\pm$ are integration constants,
from where 
\be                     \label{A4}
\beta_{\pm}=\beta_{\pm}(t_0)+\sigma_{\pm}\int_{t_0}^te^{-3\Omega}dt\,. 
\ee
The equations then reduce to 
\be               \label{A6}
\left(e^{3\Omega}\,{\dot{\Omega}}\right)^2=
\sigma^2
+\frac{1}{3}e^{6\Omega}\rho\,,
\ee
where $\sigma^2=\sigma_{-}^2+\sigma_{+}^2$ 
and $\rho$ is determined by 
the energy conservation condition,
\bea               \label{A5c}
\rho=\rho_0 e^{-3(1+w)\Omega}\,,
\eea
assuming the equation of state $P=w\rho$.

\subsubsection{Vacuum solutions}
Setting in the above formulas $\rho=\rho_0=0$,
one can integrate Eqs.\eqref{A4},\eqref{A6} to obtain 
\be                        \label{ppp0}
\alpha_a\propto   t^{p_a}\,,~~~~a=1,2,3
\ee
where 
\bea                              \label{ppp}
p_1&=&\frac13(1+\frac{\sigma_{+}+\sqrt{3}\sigma_{-}}{\sigma}),~~~
p_2=\frac13(1+\frac{\sigma_{+}-\sqrt{3}\sigma_{-}}{\sigma}),~~\notag \\
p_3&=&\frac13(1-2\frac{\sigma_{+}}{\sigma}),
\eea
so that 
\be                                  \label{AKasner}
p_1+p_2+p_3=p_1^2+p_2^2+p_3^2=1.
\ee
This corresponds to the Kasner solution. 

\subsubsection{Cosmological constant}
Let us choose in \eqref{A5c}
$w=-1$ and $\rho=\rho_0=3H^2$.
 The solution of \eqref{A6} is then   
\be                    \label{A8}
e^{3\Omega}=e^{3H(t-t_0) }-\frac{\sigma^2e^{-3H(t-t_0)}}{4H^2}.
\ee 
Inserting this to \eqref{A4} determines $\beta_{\pm}$. 
If $\sigma_\pm=0$ then the solution is pure de Sitter, $\Omega=H(t-t_0)$, 
with constant anisotropy parameters $\beta_\pm$ which can be set to zero 
via rescaling the spatial coordinates $x^a$. If $\sigma_\pm\neq 0$, then 
the solution approaches the de Sitter metric exponentially fast,  
\be
\Omega=H(t-t_0) +O(e^{-3H t }),~~~~
\beta_{\pm}=\beta_{\pm}(\infty)+O(e^{-3Ht})\,.
\ee 
Calculating the integral in \eqref{A4} gives explicitly 
\be
\alpha_a\propto e^{Ht}X_{+}^{2/3}\left(\frac{X_{-}}{X_{+}}\right)^{p_a}\,,~~~~
a=1,2,3,
\ee
where $p_a$ are the same as in \eqref{ppp} and 
\be 
X_{\pm}=1\pm\frac{\sigma}{2H}\,e^{-3H(t-t_0) }. 
\ee 
One has $X_\pm\to 1$ as $t\to\infty$, so that solutions approach de Sitter metric. 
On the other hand, taking the limit $H\to 0$ and choosing the integration 
constant $t_0$
such that $\sigma/(2H)e^{3Ht_0}=1$ one has  $X_{+}\to 2$ and 
$X_{-}\to 3Ht$ in this limit, and therefore $\alpha_a\propto   t^{p_a}$, so that 
 the Kasner solution is recovered. 

One can similarly obtain solutions for a negative cosmological constant, when 
$w=-1$ and $\rho=\rho_0=-3H^2$.
The solution of \eqref{A6} is then   
\be                    \label{A8a}
e^{3\Omega}=\frac{\sigma}{H}\,\sin(3H(t-t_0))\,,
\ee
inserting which to \eqref{A4} gives
\be
\alpha_a\propto \left(\cos\frac{3H(t-t_0)}{2}\right)^{2/3}
\left(\tan\frac{3H(t-t_0)}{2}\right)^{p_a}\,,
\ee
where $p_a$ are again the same as in \eqref{ppp}.

\subsubsection{More general matter}

For a general equation of state
Eq.\eqref{A6} reduces to 
\be                         \label{A5a}
\left(e^{3\Omega}\,{\dot{\Omega}}\right)^2=
\sigma^2
+\frac{\rho_0}{3}e^{3\Omega(1-w)}\,,
\ee
which shows that the anisotropy contribution always becomes small
for large $\Omega$, provided that $w<1$. 
This equation can be integrated in quadratures, 
the late time behavior 
of the solution is   
\be
e^{3\Omega}\sim t^{\frac{2}{w+1}},~~~~
\beta=\beta(\infty)+O(t^{\frac{w-1}{w+1}} ).
\ee
One can also consider a more general matter consisting of 
several components with different equations of state, $P_i=w_i\rho_i$,
in which case  
\bea               \label{A5}
\rho=\sum_i \rho_{i0} e^{-3(1+w_i)\Omega}\,.
\eea
For example, if there is a cosmological constant plus 
a one-component perfect fluid, 
then Eq.\eqref{A6} reduces to 
\be                         \label{A5b}
\left(e^{3\Omega}\,{\dot{\Omega}}\right)^2=
\sigma^2
+H^2 e^{6\Omega}
+\frac{\rho_0}{3}e^{3\Omega(1-w)}\,.
\ee
If $w>-1$ then for large $\Omega$
the second term on the right becomes dominant and the universe
approaches the de Sitter state.  
The conclusion is that in all cases the anisotropy effect
soon becomes negligible when the universe expands.


\section{Dynamical system description}
The second order equations \eqref{e1},  
\eqref{e2}, \eqref{e3}, \eqref{e4}
can be represented in the first order form 
\be                         \label{dyn1}
\dot{y}_N=F_N(\alpha,{\cal A},y_M),
\ee
where the variables $y_N(t)$ are defined in \eqref{vars}, while 
\begin{widetext}
\bea                            \label{dyn1a}
F_0&=&\alpha\,\frac{y_6}{y_0^2}\,, ~~~
F_1=\alpha\,\frac{y_1 y_7}{y_0^3}\,, ~~~
F_2=\sqrt{3}\,\alpha\,\frac{y_2 y_8}{y_0^3}\,, ~~~
F_3= 
{\cal A}\,\frac{y_9}{y_3^2}\,, ~~~~
F_4={\cal A}\,\frac{y_4 y_{10}}{y_3^3}\,, ~~~
F_5=\sqrt{3}\,{\cal A}\,\frac{y_5 y_{11}}{y_3^3}, \notag\\
F_6&=&{1\over 6}\left[\cos^2\eta\left(y_0\frac{\partial U}{\partial y_0}
+3\alpha U_g\right)
+3 {\alpha} \,y_0^3\,(\rho_g-P_g)
-2{\alpha}\,y_0^3\,\R\right]\,, ~~~
F_7= {1\over 12}\left[
-2\,{\cos^2\eta}\, y_1\frac{\partial U}{\partial y_1}
+{\alpha}\,y_0^3\,y_1
\frac{\partial \R}{\partial y_1}\right]\,, ~~~
~~~~\notag\\
F_8&=&{\sqrt{3}\over 12}\left[
-2\,{\cos^2\eta}\, y_2\frac{\partial U}{\partial y_2}
+{\alpha}\,y_0^3\,y_2
\frac{\partial \R}{\partial y_2}
\right]\,, ~~~
F_9={1\over 6}\left[
{\sin^2\eta}\,\left(y_3\frac{\partial U}{\partial y_3}
+3{\cal A}\, {\cal U}_f\right)
+3{\cal A}\,y_3^3\,(\rho_f-P_f) 
-2{\cal A}\,y_3^3\,\Rcal\right]\,, ~~~\notag \\
F_{10}&=&{1\over 12}\left[-2\sin^2\eta\, y_4\frac{\partial U}{\partial y_4}
+{\cal A}\,y_3^3\,y_4
\frac{\partial \R}{\partial y_4}\right]
\,, ~~~
F_{11}={\sqrt{3}\over 12}\left[
-2\,{\sin^2\eta}\, y_5\frac{\partial U}{\partial y_5}
+{\cal A}\,y_3^3\,y_5
\frac{\partial \R}{\partial y_5}
\right]\,.\notag 
\eea
Here 
$U=\alpha\, U_g+{\cal A}\,{\cal U}_f$ with  
\bea                                       \label{Va}
U_g&=&y_0^3\{b_0+b_1(\lambda_1+\lambda_2+\lambda_3) +
b_2(\lambda_1\lambda_2+\lambda_1\lambda_3+\lambda_2\lambda_3)
+b_3\lambda_1\lambda_2\lambda_3\}   \notag 
\eea
and ${\cal U}_f$ is obtained from $U_g$ by replacing $b_k\to b_{k+1}$,
\end{widetext}
where the eigenvalues $\lambda_i$ are given by
\be
\lambda_1=\frac{y_3 y_4 y_5}{y_0 y_1 y_2},~~~~~~
\lambda_2=\frac{y_3 y_4 y_2}{y_0 y_1 y_5},~~~~~
\lambda_3=\frac{y_3 y_1^2}{y_0 y_4^2}.
\ee
The 3-curvatures are 
\begin{eqnarray*}
\R=\frac{2\,n^{(1)}n^{(3)}\,y_2^2}{y_0^2y_1^2}
-\frac{1}{2}
\left(\frac{n^{(1)}\, y_2^4 y_1^6-n^{(2)}\,y_1^6
+n^{(3)}\,y_2^2}{y_0 y_2^2 y_1^4}\right)^2\,,~~
\end{eqnarray*}
while $\Rcal$ is obtained by replacing in this expression 
$y_0\to y_{3}$,  $y_1\to y_{4}$, $y_2\to y_{5}$. 
Finally, the matter terms are 
\be                              \label{mat1a}
\rho_g=\sum_i \rho_{g}^{(i)}y_0^{-3(1+w^{(i)}_g)},~~~
P_g=\sum_i w^{(i)}_g\rho_{g}^{(i)}y_0^{-3(1+w^{(i)}_g)},~~~
\ee
where $\rho_{g}^{(i)}$, $w^{(i)}_g$ 
are constant parameters, while $\rho_f$, $P_f$ are obtained 
by replacing in these expressions $y_0\to y_3$ and $\rho_{g}^{(i)}\to \rho_{f}^{(i)}$,
$w^{(i)}_g\to w^{(i)}_f$ (the number of matter components needs not 
to be the same in both sectors).

\bibliographystyle{apsrev4-1}

%

\end{document}